\documentclass[journal=jacsat,manuscript=article]{achemso}

\usepackage[utf8]{inputenc}
\usepackage[T1]{fontenc}
\usepackage{lmodern}
\usepackage{graphicx}
\usepackage{booktabs}
\usepackage{longtable}
\usepackage{array}
\usepackage{calc}
\usepackage{tabularx}
\usepackage{pdflscape}
\usepackage{placeins}
\usepackage{caption}
\usepackage[version=4]{mhchem}
\usepackage{siunitx}
\DeclareSIUnit{\angstrom}{\text{\AA}}
\usepackage{amsmath,amssymb}
\usepackage[colorlinks=true,linkcolor=blue,citecolor=blue,urlcolor=blue]{hyperref}

\setkeys{acs}{articletitle=true}
\title{Robust AI-Driven Discovery of Electronic Metal Phosphide Semiconductors}

\author{Benhao Zhu}
\altaffiliation{These authors contributed equally to this work}
\affiliation{State Key Laboratory of Integrated Optoelectronics, Key Laboratory of Automobile Materials and Key Laboratory of Material Simulation Methods and Software of MOE, School of Materials Science and Engineering, Jilin University, Changchun, Jilin 130012, China}
\author{Muhammad Faizan}
\altaffiliation{These authors contributed equally to this work}
\affiliation{State Key Laboratory of Integrated Optoelectronics, Key Laboratory of Automobile Materials and Key Laboratory of Material Simulation Methods and Software of MOE, School of Materials Science and Engineering, Jilin University, Changchun, Jilin 130012, China}
\author{Zewei Li}
\affiliation{State Key Laboratory of Integrated Optoelectronics, Key Laboratory of Automobile Materials and Key Laboratory of Material Simulation Methods and Software of MOE, School of Materials Science and Engineering, Jilin University, Changchun, Jilin 130012, China}
\author{Wenshuo Li}
\affiliation{State Key Laboratory of Integrated Optoelectronics, Key Laboratory of Automobile Materials and Key Laboratory of Material Simulation Methods and Software of MOE, School of Materials Science and Engineering, Jilin University, Changchun, Jilin 130012, China}
\author{Feifei Ren}
\affiliation{State Key Laboratory of Integrated Optoelectronics, Key Laboratory of Automobile Materials and Key Laboratory of Material Simulation Methods and Software of MOE, School of Materials Science and Engineering, Jilin University, Changchun, Jilin 130012, China}
\author{Jiahao Xie}
\affiliation{State Key Laboratory of Integrated Optoelectronics, Key Laboratory of Automobile Materials and Key Laboratory of Material Simulation Methods and Software of MOE, School of Materials Science and Engineering, Jilin University, Changchun, Jilin 130012, China}
\email{xiejh.mail@gmail.com}
\author{Lijun Zhang}
\affiliation{State Key Laboratory of Integrated Optoelectronics, Key Laboratory of Automobile Materials and Key Laboratory of Material Simulation Methods and Software of MOE, School of Materials Science and Engineering, Jilin University, Changchun, Jilin 130012, China}
\email{lijun_zhang@jlu.edu.cn}


\begin{document}

\begin{abstract}

Metal phosphides have diverse bonding motifs and coordination environments, making them promising for optoelectronic and thermoelectric applications, but their chemical space remains underexplored. Here we report an AI-driven high-throughput discovery workflow that combines generative materials design, machine-learning interatomic potentials, and targeted density functional theory (DFT) calculations. ICSD-derived Wyckoff-site substitution and MatterGen-based conditional structure generation are used to expand the candidate space beyond existing phosphide databases. A domain-finetuned DPA3 machine-learning potential then enables efficient prescreening of thermodynamic and dynamical stability before DFT validation. This workflow identifies 3,574 previously unreported stable phosphide structures, including 196 semiconductors with HSE06 band gaps of 0-3.0 eV. By screening these new semiconductors together with experimentally known phosphide semiconductors, we identify 30 promising optoelectronic candidates and 26 promising thermoelectric candidates, including seven newly discovered optoelectronic materials and eight newly discovered thermoelectric materials. These results provide a candidate pool for experimental synthesis and show that combining generative AI with machine-learning interatomic potentials can accelerate the discovery of functional semiconductor materials.

\end{abstract}

\section{Introduction}

Metal phosphides constitute an important family of inorganic materials characterized by rich structural chemistry and a wide range of tunable physical properties. Owing to these characteristics, they have attracted considerable attention in the search for next-generation functional semiconductors.\cite{ref01,ref02,ref03} In several subclasses, the interplay between ionic interactions among cations and covalent polyanionic frameworks gives rise to their semiconductor behavior.\cite{ref04,ref05,ref06} Moreover, their rich compositional space, varied local coordination environments, and structural diversity provide substantial opportunities for tailoring band structures and transport properties.\cite{ref07,ref08} This tunability, arising from the interplay between bonding characteristics and local atomic arrangement, makes metal phosphides particularly promising for both optoelectronic and thermoelectric applications. Several Zintl phosphides and triel-phosphide systems have already demonstrated optimum band gaps, strong absorption in the visible range, and high defect tolerance suitable for photovoltaics,\cite{ref02,ref09,ref10,ref11,ref12} while many phosphides also exhibit high carrier mobility, low lattice thermal conductivity, and excellent thermoelectric performance.\cite{ref13,ref14,ref15,ref16}

Despite this promise, the computational discovery of previously unknown metal phosphides remains challenging. Over the past decade, database-driven high-throughput screening has significantly improved the efficiency of exploring known materials spaces and has identified a series of promising phosphide candidates for energy-conversion applications.\cite{ref17} However, such approaches are inherently limited by existing crystal-structure databases and therefore offer limited access to structure types that have not yet been reported. To extend the search space, researchers have employed prototype-substitution and local-structure-enumeration strategies. These methods effectively expand candidate pools within the neighborhood of known topologies and have facilitated the discovery of several functional phases.\cite{ref03,ref18,ref19,ref20,ref21,ref22} Nevertheless, these approaches also have important limitations. Since they rely on predefined structural templates, they cannot readily access structure spaces defined by entirely new coordination environments and bonding motifs. Crystal structure prediction methods, such as CALYPSO and USPEX, can, in principle, identify low-energy configurations from scratch for a given composition and have successfully predicted several simple binary phosphides.\cite{ref23,ref24} However, metal phosphides exhibit significant polymorphism because of the flexibility of polyanionic phosphorus frameworks. A single stoichiometry may therefore support multiple competing configurations with similar energies but distinct local environments. In such complex phase spaces, the computational cost of global structure searches scales rapidly with compositional complexity and atom size, limiting their applicability in large-scale exploration. Consequently, moving beyond known prototypes while keeping the computational cost tractable is a central challenge in the discovery of unknown metal-phosphide semiconductors.

Generative artificial intelligence (AI) has recently emerged as a promising route for inorganic-crystal discovery, building on broader advances in high-throughput computation, data-driven optimization, and machine-learning-assisted screening for functional semiconductors.\cite{ref25,ref26,ref27,ref28,ref29} Unlike conventional enumeration and prototype-based extrapolation strategies, generative models learn structural distributions from existing crystal datasets and can generate new candidates under constraints related to composition, crystal symmetry, and target properties\cite{ref30}. In this way, the search can expand from the vicinity of known structures into a far broader unexplored materials space. Previous studies have already shown the potential of this strategy in semiconductor screening and topological-material discovery.\cite{ref31,ref32,ref33,ref34} However, expanding the candidate space alone does not eliminate the computational bottleneck. For large numbers of candidate structures, full DFT relaxation and stability calculations remain prohibitively expensive. Universal machine-learning interatomic potentials (MLIPs) have therefore emerged as a key bridge between candidate generation and first-principles validation. M3GNet has been employed for rapid relaxation and stability assessment of millions of hypothetical crystals,\cite{ref35} while DPA3 and UMA further offer stronger cross-system generalization and higher computational efficiency.\cite{ref36,ref37} More importantly, these models are increasingly being integrated into AI-assisted materials-discovery workflows and have already shown considerable potential in accelerating closed loop frameworks linking theoretical prediction and experimental validation.\cite{ref38,ref39} For metal phosphides, where local chemical environments are highly complex, phosphorus bonding motifs are diverse, and candidate space is vast, a hierarchical AI-driven strategy that combines generative models with universal MLIPs is particularly attractive. Generative models expand the search beyond prototype limitations, whereas MLIPs enable low-cost, large-scale stability screening before DFT calculations are reserved for precise validation of the most promising candidates.

Against this background, in this work, we develop an AI-driven hierarchical framework for discovering unknown metal-phosphide semiconductors that combine dual-route candidate generation with staged stability and property screening. ICSD-derived Wyckoff substitution is used to systematically extend known prototype families, while MatterGen-based conditional generation is used to access structures beyond existing prototypes. This approach expands the candidate space to approximately 137,000 configurations spanning 76 elements. Subsequently, the DPA3 universal deep-learning potential, fine-tuned for phosphide local chemical environments, is employed to prescreen thermodynamic and dynamical stability, thereby focusing expensive DFT calculations on the most promising candidates for detailed electronic-structure and functional-property evaluation. Using this workflow, we identify 196 previously unreported semiconductors. Through a combined screening of both newly discovered and experimentally known phosphide semiconductors, we further identify 30 promising optoelectronic and 26 promising thermoelectric candidates, including seven newly discovered optoelectronic materials and eight newly discovered thermoelectric materials. These findings demonstrate that the proposed AI-driven hierarchical framework significantly expands the accessible structural space of unknown metal phosphides and provides a transferable strategy for discovering underexplored semiconductor systems.

\section{Results and Discussion}

\subsection{Dual-Track Generative Materials Design}

\begin{figure}[!p]
\centering
\includegraphics[height=0.64\textheight,width=\linewidth,keepaspectratio]{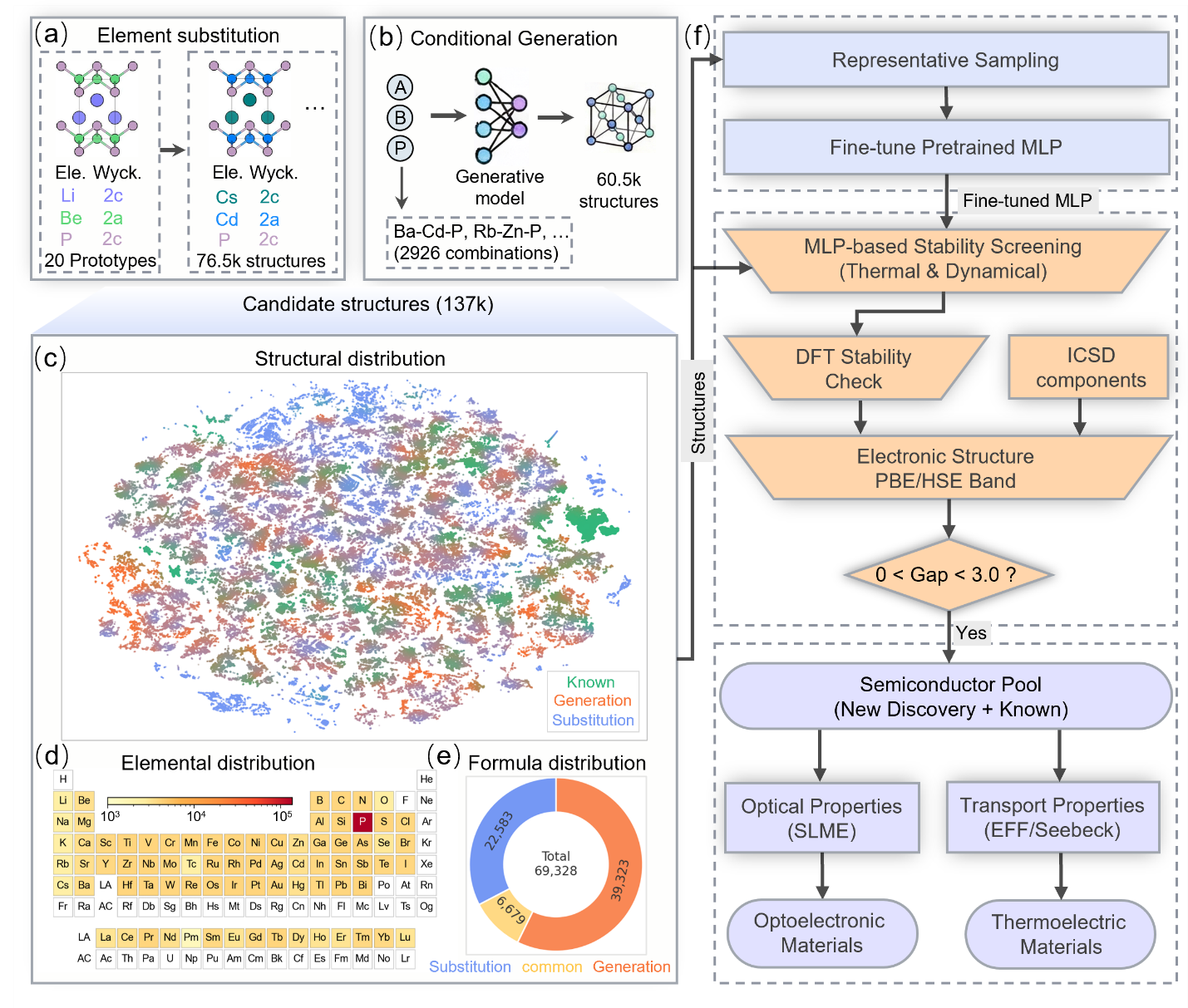}
\caption{Overall workflow of the generative high-throughput discovery framework developed for metal phosphide semiconductors. (a) An element-substitution route based on known crystal prototypes, used to expand candidate compositional space while preserving key crystallographic backbone features. (b) A conditional structure-generation route based on a generative machine-learning model, used to explore structures beyond existing databases under prescribed compositional and stability constraints. (c-e) Coverage of the candidate pool in structural, elemental, and compositional space. (c) Structural-space distribution of the candidate pool, including known metal phosphides and structures generated by the machine-learning and element-substitution routes; green, orange, and blue denote known, generative-model-derived, and substitution-derived structures, respectively. (d) Element-wise count distribution of candidate structures. (e) Comparison of the numbers of distinct chemical formulas contributed by different discovery routes after excluding previously reported metal phosphide formulas. (f) End-to-end high-throughput screening workflow, spanning candidate generation, thermodynamic and dynamical stability screening, and functional property evaluation.}
\label{fig:workflow}
\end{figure}

To overcome the limited diversity of direct database retrieval and the structural bias associated with a single prototype-enumeration route, we developed a dual-track AI-driven discovery framework for metal phosphides, as shown in Figure~\ref{fig:workflow}. The framework combines prototype substitution with generative modeling (Figure~\ref{fig:workflow}a,b), together with MLP-accelerated prescreening, DFT validation, and property-oriented selection. The two generation routes play complementary roles in exploring the phosphide chemical space. The substitution route preserves crystallographic backbones inherited from known prototypes and therefore efficiently expands the neighborhood of experimentally accessible structure families. In contrast, the generative-model route moves beyond existing structural templates, enabling exploration of candidate configurations that may not be represented in current databases. Within this framework, the prototype-substitution route generated approximately 76,500 candidate structures, and the generative-model route contributed an additional \textasciitilde{} 60,500 structures (see Computational Methods for more details).

The resulting candidate pool exhibits broad coverage across structural, elemental, and compositional space. In the structural embedding (shown in Figure~\ref{fig:workflow}c), the generated candidates not only overlap with regions occupied by known ICSD phosphides but also extend into sparsely populated and previously unexplored areas, indicating that the workflow does not merely reproduce database-like structures. At the elemental level (Figure~\ref{fig:workflow}d), broad participation across the periodic table demonstrates that the generated structures sample a chemically diverse space rather than being confined to a narrow set of compositions. At the compositional level (Figure~\ref{fig:workflow}e), after excluding formulas already reported for metal phosphides, the two discovery routes together contribute more than 60,000 distinct new chemical formulas. Notably, these new formulas include both variants related to known prototype families and candidates that are not readily traceable to existing database entries. Taken together, these results show that the dual-track strategy expands the searchable phosphide space along two complementary directions: it systematically enlarges the local neighborhood surrounding known compounds while simultaneously opening access to genuinely unexplored regions of chemical and structural space.

\FloatBarrier

\subsection{Descriptor-driven DPA3 fine-tuning for high-fidelity screening}

\begin{figure}[!htbp]
\centering
\includegraphics[height=0.34\textheight,width=\linewidth,keepaspectratio]{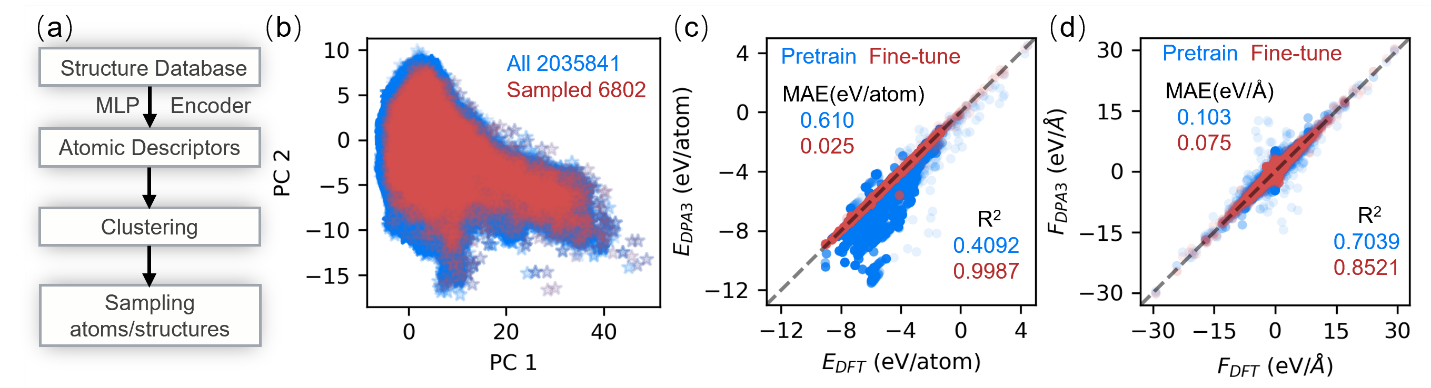}
\caption{Sampling strategy and fine-tuning performance of the DPA3 machine-learning potential. (a) Workflow for dataset based on featurization, clustering, and sampling of the local atomic environment. (b) Coverage of the candidate-structure space by the sampled data. (c) Comparison of energy predictions on the test set before and after fine-tuning. (d) Comparison of atomic-force predictions on the test set before and after finetuning.}
\label{fig:dpa3}
\end{figure}

Once the large candidate pool had been established, the key challenge in prescreening stability was to build a surrogate model capable of accurately describing the complex local environments present in metal phosphides. Figure~\ref{fig:dpa3} shows that the pretrained DPA3 potential exhibits errors that are too large for direct deployment in the present system, especially because reliable stability screening requires consistent accuracy across a broad range of compositions and structural motifs. The candidate pool spans a wide range of valence states, coordination environments, local distortions, and lattice stiffnesses. Consequently,a randomly assembled finetuning dataset would be unlikely to provide balanced coverage of the environments most relevant to large-scale screening. To address this, we constructed the finetuning dataset using a descriptors-driven sampling strategy designed to maximize coverage of the local-environment descriptor space of the full candidate pool rather than sampling density alone. Specifically, more than two million local atomic environments were first encoded into descriptor space, followed by dimensionality reduction and clustering to identify representative regions for DFT labeling (see Computational Methods for more details). This yielded 6,802 representative local environments, which were then traced back to 6,018 parent structures. The two-dimensional embedding shown in Figure~\ref{fig:dpa3}b demonstrates that the selected dataset provides broad coverage of the candidate local-environment space, including several low-density regions that would be difficult to capture through conventional random selection. This is particularly important because rare but chemically distinct environments can disproportionately affect stability predictions when screening a large and chemically diverse candidate pool.

\FloatBarrier

Finetuning on this descriptor-informed dataset substantially improves agreement between the machine-learning potential and DFT calculations. The mean absolute error in energy decreases from about 620 to 25 meV/atom, while the force error is reduced from approximately 103 to \SI{75}{\milli\electronvolt\per\angstrom}. Beyond the reduction in average errors, the post-finetuning results also exhibit a much stronger correlation with DFT predictions across the entire test set, indicating a more transferable description of local bonding environments. These results demonstrate that local-environment-coverage-driven sampling is an effective strategy for adapting a general pretrained potential to the chemically diverse phosphide space, thereby providing a more reliable basis for subsequent large-scale stability screening.

\subsection{Hierarchical stability screening for metal-phosphide semiconductors}

\begin{figure}[!p]
\centering
\includegraphics[height=0.733\textheight,width=\linewidth,keepaspectratio]{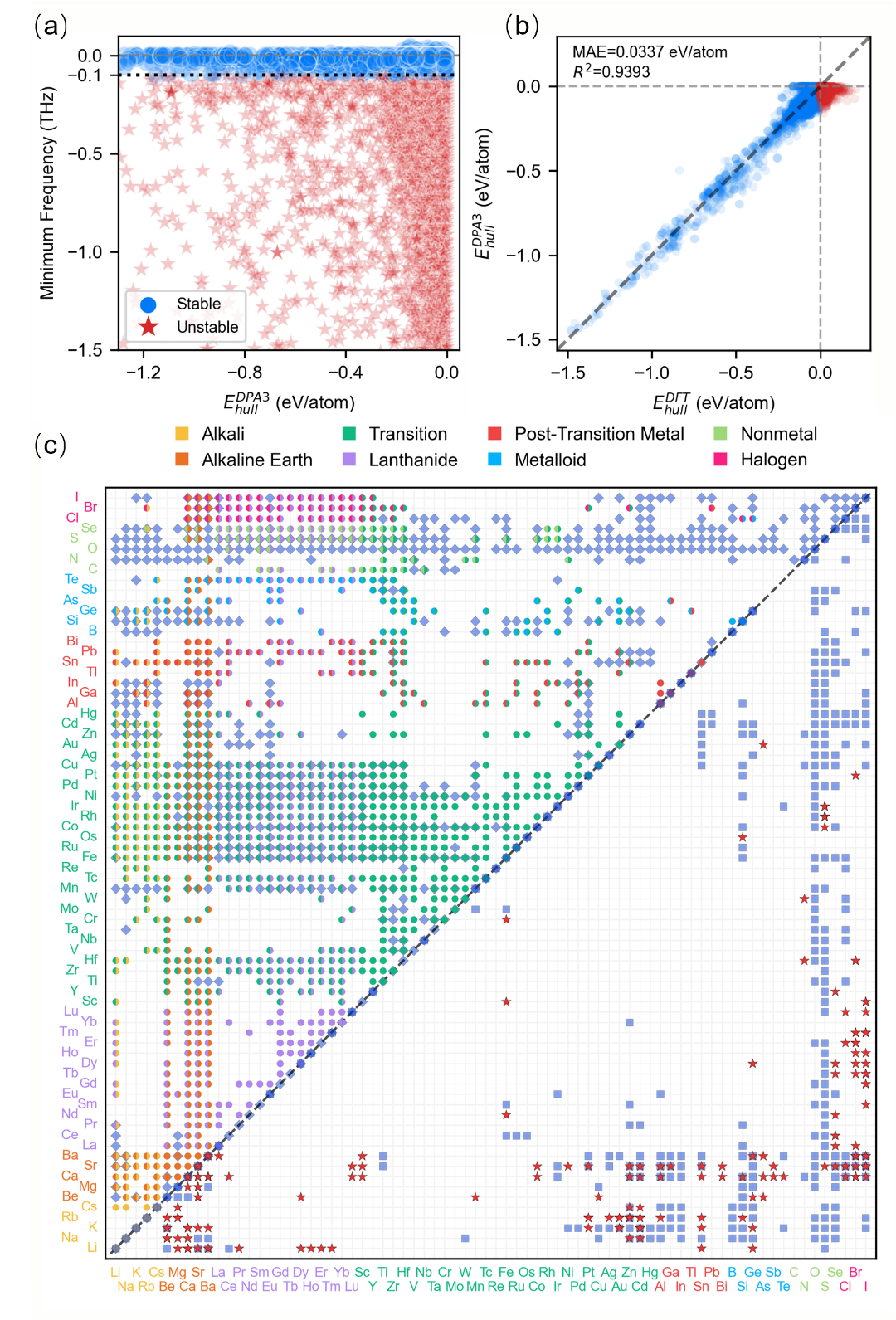}
\caption{MLP-based thermodynamic and dynamical screening, DFT validation of thermodynamic stability, and distribution of stable structures and semiconductor candidates across elemental-combination space. (a) High-throughput thermodynamic and dynamical stability screening based on the machine-learning potential. (b) Comparison between machine-learning predictions and DFT validation for thermodynamic stability. (c) Distribution of DFT-validated thermodynamically stable structures and semiconductor candidates across elemental-combination space. Blue squares in the upper-triangle denote experimentally reported phosphide materials, and those in the lower-triangle represent experimentally known semiconducting phosphides. Circles in the upper triangle represent newly designed thermodynamically stable compounds, with colors indicating the constituent elements, and pentagrams in the lower triangle mark newly identified semiconductor candidates.}
\label{fig:stability-screening}
\end{figure}

Using the finetuned DPA3 potential, we first screened the candidate structures for thermodynamic stability and identified 10,753 low-energy structures lying within 1 meV/atom of the convex hull. After applying an additional dynamical stability criterion based on the machine learning potential, 6,067 structures remained. Dynamical stability defined by a minimum phonon frequency no lower than -0.1 THz, a threshold chosen to accommodate small numerical imaginary modes arising from finite-displacement phonon calculations and MLIP approximations (Figure~\ref{fig:stability-screening}a). Subsequent DFT structural relaxation and re-calculation of formation energies relative to the DFT convex hull yielded 4,042 thermodynamically stable structures at the DFT level (Figure~\ref{fig:stability-screening}b). Among these, 468 overlap with experimentally reported compounds, whereas the remaining 3,574 are previously unreported stable structures retained for subsequent screening. The accuracy of the machine-learning screening is confirmed by the excellent agreement between machine-learning and DFT hull energies, with a mean absolute error of only 33.7 meV/atom. Furthermore, supplementary Figure~\ref{fig:workflow}a,b shows a Spearman correlation coefficient of 0.9994, indicating high consistency in relative-energy ranking between the machine-learning model and DFT. Together, these results suggest that the finetuned DPA3 potential provides sufficient accuracy for large-scale thermodynamic classification and candidate prioritization. To further screen for functional properties, we next performed electronic-structure calculations to identify semiconductor candidates. Among the 3,574 previously unreported stable compounds, 252 were initially selected based on PBE band gaps between 0 and 3.0 eV. To correct for the well-known systematic underestimation of semilocal functionals,\cite{ref40,ref41} we then computed HSE06 band gaps for these candidates and ultimately identified 196 previously unreported semiconductors with HSE06 gaps in the range of 0-3.0 eV. In addition, 148 experimentally reported phosphides fall within the same HSE06 bandgap window, yielding a total of 344 semiconducting phosphides for subsequent property analysis.

To further understand the chemical factors associated with thermodynamic stability and semiconductor formation in phosphorus-containing compounds, we mapped both the DFT-validated stable structures and the semiconducting subset onto the elemental-combination space shown in Figure~\ref{fig:stability-screening}c. The horizontal and vertical axes represent the two non-P element classes in each composition, with the diagonal corresponding to binary phosphides. The upper and lower triangles of the symmetric matrix display thermodynamically stable and semiconducting compounds, respectively. Thermodynamically stable compounds span a broad elemental space but exhibit clear chemical clustering. These phases are enriched in transition-metal-containing phosphide systems, particularly those paired with alkali metals, alkaline-earth metals, or other transition metals, and also appear in certain halogen-containing compositions. These patterns suggest that such chemical combinations are tend to favor low-energy phosphorus-containing frameworks. To further elucidate the origins of these trends, we performed SHAP attribution analysis using a random-forest model (Supplementary Figure~\ref{fig:workflow}c,d). The results identify the local coordination environment as the dominant factor associated with phase stability, with descriptors related to sixfold coordination exhibiting relatively high feature importance. Structures containing a larger proportion of octahedral and distorted sixfold coordination generally display more negative SHAP values, indicating a strong tendency toward low formation energies. Elemental chemistry also plays a significant role. Descriptors associated with valence-electron differences, atomic-size mismatch, and periodic table group-position contrast are all closely associated with stability. Compared with the broader stable compound distribution, the semiconducting subset is distinctly more localized, concentrating in chemistries where alkali or alkaline-earth metals pair with post-transition metals, metalloids, or closed-shell elements. This distribution is consistent with Zintl-like bonding mechanism, in which electropositive cations donate electrons to the P-centered polyanionic frameworks, enabling electron-precise covalent bonding and the formation of a finite band gap. Conversely, many transition-metal-rich stable phosphides remain metallic because partially occupied d states persist near the Fermi level and hinder band-gap opening.\cite{ref04,ref42,ref43} These findings suggest that semiconductor formation in metal phosphides requires the simultaneous satisfaction of several conditions, including favorable local coordination, compatible elemental pairing, and electron-precise bonding configurations.

\FloatBarrier

\subsection{Functional property screening of phosphide semiconductors}

\begin{figure}[!htbp]
\centering
\includegraphics[height=0.46\textheight,width=\linewidth,keepaspectratio]{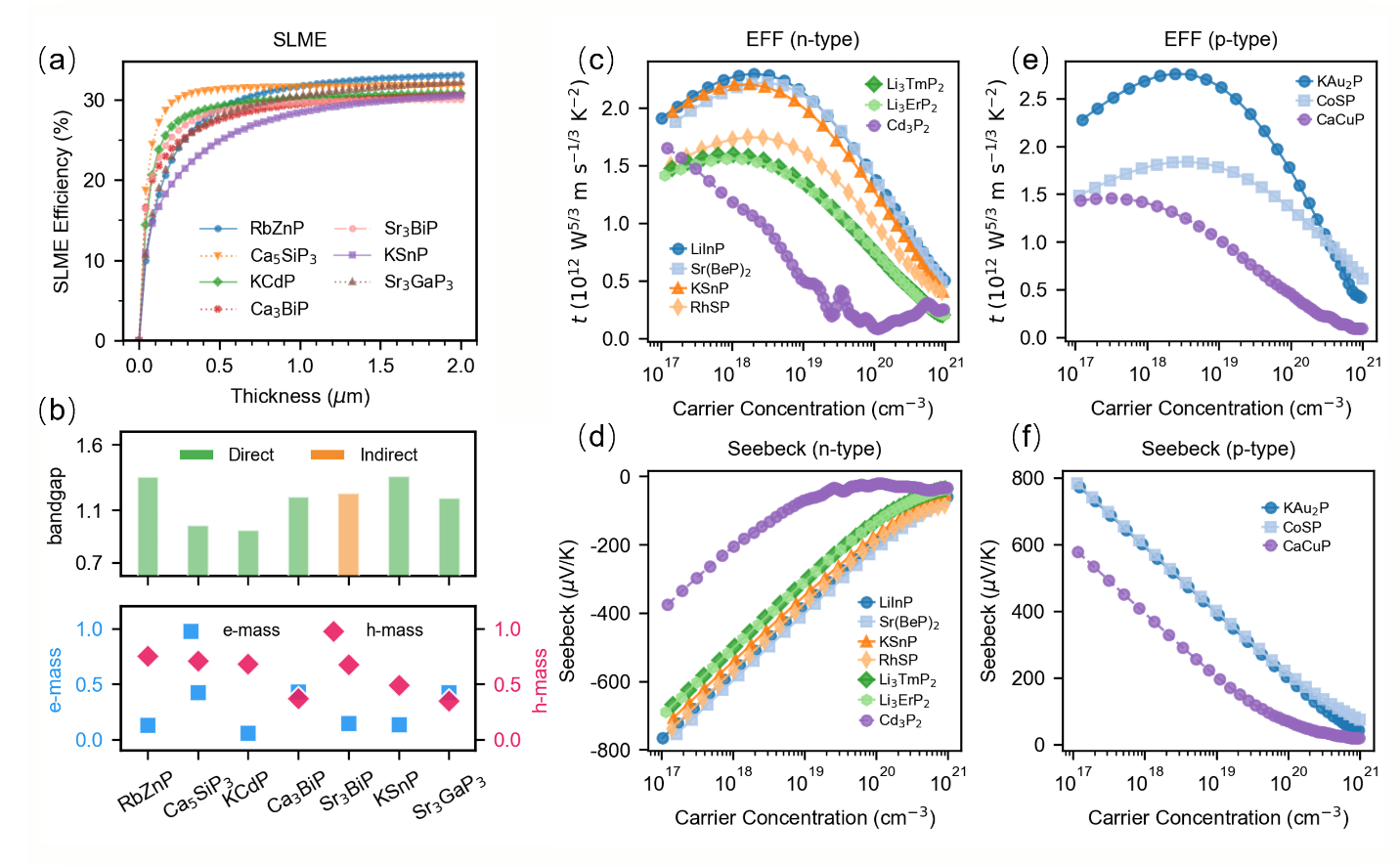}
\caption{Functional properties of stable phosphide semiconductors for optoelectronic and thermoelectric applications. (a) SLME as a function of absorber thickness for newly identified photovoltaic absorber candidates. (b) Distribution of HSE06 band gaps, electron effective masses, and hole effective masses for the selected optoelectronic candidates. (c,e) EFF as a function of carrier concentration for newly identified thermoelectric candidates under n-type and p-type conditions. (d,f) Corresponding Seebeck coefficients as a function of carrier concentration for the n-type and p-type thermoelectric candidates.}
\label{fig:functional-screening}
\end{figure}

Starting from the semiconducting phosphides identified above, including both the newly discovered compounds and experimentally reported materials, we next evaluated their potential for optoelectronic and thermoelectric applications. Although both target applications rely on semiconducting electronic structures, the key performance descriptors differ substantially. For optoelectronic screening, an ideal photovoltaic absorber should possess a suitable band gap, strong optical absorption, and favorable bipolar carrier transport. We therefore assessed photovoltaic potential using the spectroscopic limited maximum efficiency (SLME), which explicitly accounts for absorption strength and absorber thickness, and is thus more realistic than the Shockley-Queisser limit for thin-film photovoltaic materials. To identify materials with both high theoretical efficiency and sufficient thickness tolerance, materials required an SLME exceeding 30\% at an absorber thickness of \SI{2}{\micro\meter} and 25\% at \SI{0.5}{\micro\meter}. Notably, an SLME above 30\% at \SI{2}{\micro\meter} already places candidates in the efficiency range associated with leading single-junction absorbers such as GaAs.\cite{ref44} To further reduce the likelihood of transport-limited behavior, electron and hole effective masses were additionally constrained to remain below $1~m_0$. Under these combined criteria, 30 promising photovoltaic absorbers were identified, including 7 newly discovered compounds and 23 known phosphides. The SLME curves and distributions of HSE06 band gaps and carrier effective masses for the newly discovered compounds are presented in Figure~\ref{fig:functional-screening}a,b, with detailed structural, electronic, and vibrational properties provided in Supplementary Table~\ref{tab:si-2} and Supplementary Figure~\ref{fig:si-new-photoactive}. The absence of imaginary phonon modes further confirms the dynamical stability of these newly discovered candidates. The results for the known phosphides, which include established optoelectronic materials such as \ce{InP} and \ce{BaCd2P2} as well as several previously unexplored phosphides, are summarized in Supplementary Table~\ref{tab:si-3} and Supplementary Figures~\ref{fig:si-known-photoactive-slme} and~\ref{fig:si-known-photoactive-bands}. Collectively, these screening results indicate that metal phosphides contain a promising class of absorber materials that satisfy critical criteria for both SLME and carrier transport, while also highlighting substantial underappreciated optoelectronic potential within the phosphide family.

Thermoelectric screening entails distinct requirements. The central challenge lies not merely in achieving a large Seebeck coefficient, but in identifying materials that can partially overcome the intrinsic trade-off between thermopower and electrical conductivity.\cite{ref45,ref46,ref47} To this end, we employed the electronic fitness function (EFF) as the primary screening descriptor and evaluated both n-type and p-type transport. EFF has been widely adopted to identify materials with favorable thermoelectric transport potential by balancing band-structure complexity with carrier mobility.\cite{ref48,ref49,ref50} Candidates were retained when the peak EFF within the carrier-concentration range of $10^{17}$ to $10^{21}\ \mathrm{cm}^{-3}$ exceeded $2.0 \times 10^{12}\ \mathrm{W}^{5/3}\mathrm{m}\mathrm{s}^{-1/3}\mathrm{K}^{-2}$, a threshold comparable to representative thermoelectric phosphides such as \ce{Cd3P2} and \ce{CaCuP}. This screening yielded 26 promising thermoelectric materials, including 8 newly discovered compounds and 18 known phosphides as promising thermoelectric candidates. Figure~\ref{fig:functional-screening}c and~\ref{fig:functional-screening}e show the EFF as a function of carrier concentration for the newly identified candidates under the n-type and p-type candidates, respectively, while Figure~\ref{fig:functional-screening}d and~\ref{fig:functional-screening}f present the corresponding Seebeck coefficients, illustrating the evolution of thermopower across the same carrier-concentration range. These results reveal that several candidates maintain high EFF values over a broad doping window, indicating high thermoelectric performance potential. Supplementary Table 4 and Supplementary Figure~\ref{fig:si-new-thermoelectric} provide additional structural, electronic, and phonon data for the newly discovered compounds. All candidates exhibit phonon spectra likewise free of imaginary frequencies, confirming dynamical stability. The corresponding results for experimentally known phosphides, including benchmark thermoelectric materials such as \ce{Cd3P2} and \ce{CaCuP}, as well as several underexplored phosphides, are summarized in Supplementary Figures~\ref{fig:si-known-te-eff} and~\ref{fig:si-known-te-bands}. Overall, these results indicate that metal phosphides constitute a versatile set of semiconductor platforms capable of supporting both optoelectronic and thermoelectric functionalities.

\FloatBarrier

\begin{figure}[!p]
\centering
\includegraphics[height=0.76\textheight,width=\linewidth,keepaspectratio]{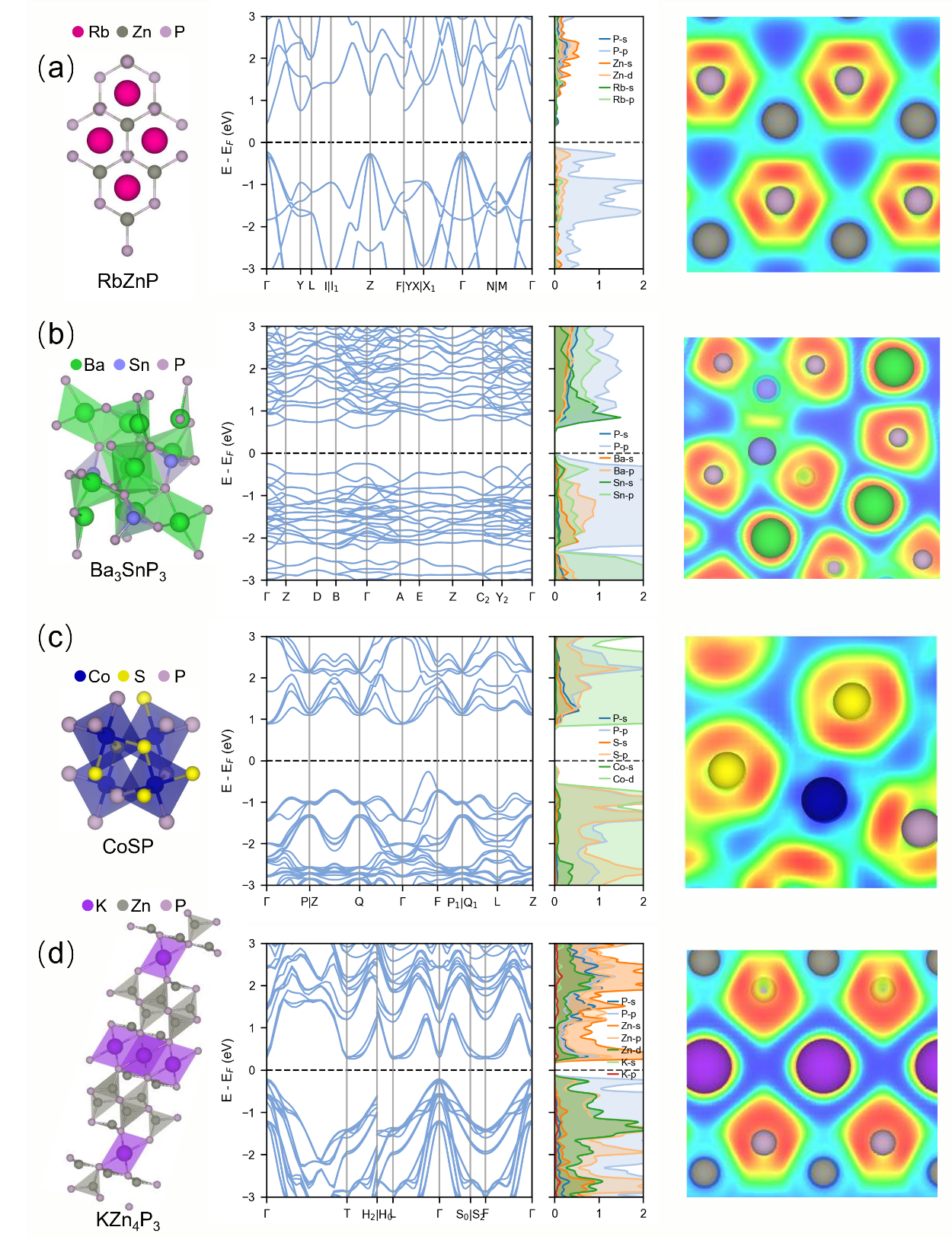}
\caption{Electronic-structure and stability analyses of representative functional semiconductor candidates. (a) Crystal structure, electronic band structure, projected density of states, and electron localization function of the representative optoelectronic candidate \ce{RbZnP}. (b) \ce{Ba3SnP3}, (c) p-type thermoelectric candidate \ce{CoSP}, and (d) n-type thermoelectric candidate \ce{KZn4P3}.}
\label{fig:electronic-structures}
\end{figure}

To elucidate the physical mechanisms underlying the predicted functional properties, we further analyzed the electronic structures of representative optoelectronic and thermoelectric candidates selected from both the newly discovered and experimentally known phosphides, as shown in Figure~\ref{fig:electronic-structures}. For optoelectronic applications, \ce{RbZnP} and \ce{Ba3SnP3} reveal a common design principle for high performance photovoltaic absorbers despite their markedly distinct structural dimensionalities. In both compounds, the band-edge states are governed primarily by covalent metal-phosphorus bonding networks rather than the electropositive cation sublattices. Such localization on electronically connected metal-phosphorus networks provides strong band-edge orbital overlap and favorable band dispersion, supporting efficient optical transitions and balanced carrier transport.\cite{ref02,ref51} As Shown in Figure~\ref{fig:electronic-structures}a, \ce{RbZnP} adopts a layered [ZnP] framework and exhibits a direct-gap semiconductor nature with an HSE06 gap of 1.36 eV at $\Gamma$ Position. Both the valence-band maximum (VBM) and conduction-band minimum (CBM) band edges are highly dispersive, indicative of favorable carrier mobility. The CMB is mainly derived from P-s and Zn-s states, whereas the VBM is dominated by P-p and Zn-d states with clear signatures of p-d hybridization. This band-edge configuration is favorable for efficient bipolar carrier transport while maintaining absorber quality.\cite{ref02,ref52} The ELF further reveals pronounced electron sharing along the Zn-P bonds, supporting that the assignment of the band-edge states to a strongly covalent [ZnP] bonding network. \ce{Ba3SnP3} (Figure~\ref{fig:electronic-structures}b) follows the same general principle within a three-dimensional Sn-P network. Although the material is formally an indirect-gap semiconductor, the difference between its indirect and direct gaps is only \textasciitilde{} 3.3 meV, giving it an effectively quasi-direct-gap character.\cite{ref53} More importantly, both the CBM and VBM are mainly associated with the Sn-P bonding network. The conduction edge is dominated by Sn-s and P-p states, while the valence edge is primarily composed of P-p and Sn-p orbitals. The ELF is fully consistent with this interpretation, revealing predominantly ionic Ba-P interactions but significantly stronger electron localization along the Sn-P bonds. Although \ce{RbZnP} and \ce{Ba3SnP3} have different structural dimensionalities, they together suggest that favorable optoelectronic performance in phosphides is governed primarily by the electronic character of the band-edge states. In both compounds, these states are controlled by covalent metal-phosphorus bonding networks, which support strong optical transitions, favorable band dispersion, and balanced carrier transport.

For thermoelectric applications, \ce{CoSP} and \ce{KZn4P3} demonstrate two distinct band-edge routes toward strong transport performance. In p-type \ce{CoSP} (Figure~\ref{fig:electronic-structures}c), the key feature is not simply the VBM itself but the overall structure of the upper valence manifold. The highest valence band is highly dispersive, which is favorable for hole mobility, while a pronounced high-density-of-states manifold appears approximately 0.4-0.5 eV below the VBM. This deeper manifold originates mainly from S-p and Co-d orbital degeneracy, consistent with the localized Co-S bonding indicated by the ELF, and is associated with multiple band crossings and local extrema along the Z-Q path and near the F point. Consequently, a large number of hole states become accessible within a relatively narrow energy window. Because the region between the VBM and the deeper manifold is occupied by a comparatively low-density-of-states dispersive band, the Fermi level can be shifted into the high-density-of-states region without requiring excessively large carrier concentrations. \ce{CoSP} therefore combines the mobility advantage of a dispersive band edge with the enhanced thermopower associated with degenerate valence manifold and high density of states, thereby accounting for its simultaneously high Seebeck coefficient and EFF.\cite{ref54,ref55} In contrast, the thermoelectric performance of \ce{KZn4P3} (Figure~\ref{fig:electronic-structures}d) exhibits an n-type mechanism governed primarily by conduction-band engineering. The CBM, composed mainly of Zn-s, Zn-p, and P-p states within the covalent Zn-P network indicated by the ELF, contains multiple closely spaced local minima distributed near the , T, L, and F points. This produces a multi-valley conduction edge with significant valley degeneracy. Such a band-edge configuration enhances the density-of-states effective mass through both valley multiplicity and band degeneracy while retaining efficient electron transport, thereby supporting strong n-type thermoelectric performance.\cite{ref54,ref56} Taken together, these representative examples show that promising thermoelectric performance in phosphides can be achieved through different forms of band-edge complexity, including doping-accessible valence-band convergence for p-type transport and intrinsic multi-valley conduction architectures for n-type transport.

In addition to intrinsic functional performance, practical deployment also depends on economic viability. We therefore estimated the compositional cost of the screened functional candidates and analyzed the relationship between material cost and predicted performance, with known materials included for comparison (Supplementary Figure~\ref{fig:si-cost-performance}). For optoelectronic materials, the cost-SLME map shows several newly discovered candidates fall in a competitive region combining high predicted efficiencies with relatively low material cost. For thermoelectric materials, the cost-performance distribution based on peak EFF identifies several candidates with promising transport potential across a broad range of compositional costs. Overall, these results suggest that phosphorus-containing semiconductors are attractive not only from a performance perspective but also in terms of performance-to-cost balance, further enhancing their prospects for practical energy-conversion applications.

\clearpage
\FloatBarrier
\clearpage

\section{Conclusions}

In summary, we developed an AI-driven hierarchical generative discovery framework for metal-phosphide semiconductors that links generative candidate design, large-scale stability screening, first-principles validation, and property-oriented selection within a unified workflow. By combining ICSD-derived Wyckoff substitution with MatterGen based conditional crystal generation, we explored 2,926 elemental combinations, generated approximately 137,000 candidate structures, and expanded the accessible search space to more than 60,000 previously unexplored chemical formulas. Following descriptor-coverage-driven fine-tuning, the DPA3 machine learning potential achieved a substantial improvement in predictive accuracy, reducing the energy error from \textasciitilde{} 620 to \textasciitilde{} 25 meV/atom and enabling a reliable surrogate model for thermodynamic and dynamical stability screening. Using this workflow, we identified 3,574 previously unreported DFT-stable phosphide structures, including 196 previously unreported semiconductors. Subsequent functional screening further identified seven promising optoelectronic and eight promising thermoelectric candidates with excellent predicted performance.

Beyond accelerating materials discovery, this work provides insight into the fundamental factors governing stability and semiconductor formation in metal phosphides. Interpretable machine-learning analyses indicate that near-octahedral local coordination, chemically compatible elemental combinations, and electron-precise bonding are key ingredients in stabilizing low-energy phosphide frameworks. Together, the resulting candidate materials establish a valuable foundation for experimental synthesis and validation, while the hierarchical integration of generative AI, machine-learning interatomic potentials, and first-principles calculations offers a transferable paradigm for accelerating the discovery of functional materials in other underexplored materials families.

\section{Computational Methods}

\subsection{Generation of candidate structures}

Candidate metal-phosphide semiconductors were constructed through two complementary routes. In the first route, phosphorus-containing inorganic structures were screened from the ICSD database,\cite{ref57} excluding entries containing O, F, noble gases, or actinides, which yielded a total of 1,279 structures. Structural symmetry was then analyzed using the SpacegroupAnalyzer module in pymatgen to extract Wyckoff letters and multiplicities and to construct element-agnostic Wyckoff-pattern labels for prototype identification. Based on a combined consideration of prototype occurrence frequency, structural diversity, and the number of atoms in the unit cell, 20 representative prototypes were manually selected for subsequent elemental substitution (Supplementary Table~\ref{tab:si-1}). For each selected prototype, a generic XYP stoichiometry was adopted, and elements from Li to Bi, excluding F, noble gases, and actinides, were systematically substituted while preserving the parent space-group symmetry and Wyckoff occupations. In parallel, a pretrained MatterGen chemical\_system\_energy\_above\_hull model was used to directly generate new crystal structures under constraints on elemental composition and hull-related stability. Based on pairwise combinations of 76 elements, 2,926 unique elemental combinations were generated and subsequently combined with P to form ternary systems. Each chemical system was sampled once to generate 24 structures with a guidance strength of 0.1, after which structures incompatible with the designated elemental combinations were discarded. Together, these two routes produced the candidate pool for subsequent stability and property screening.

\subsection{Fine-tuning of phosphide-specific machine-learning potential}

To enable efficient large-scale stability screening, a phosphide-specific machine-learning potential was constructed by fine-tuning the pretrained DPA3 model. The fine-tuning labels were obtained from PBE-level DFT calculations. Training structures were selected using DIRECT (DImensionality-Reduced Encoded Clusters with sTratified sampling).\cite{ref58} Unlike the standard workflow, which samples directly from structure descriptors, we first reduced and clustered descriptors associated with local atomic environments from the full candidate pool and then mapped representative local environments back to their parent structures. This strategy enabled the construction of a fine-tuning dataset better aligned with phosphide chemistry. Birch clustering was employed with threshold\_init = 0.16. The final dataset was randomly split into training and validation subsets at a ratio of 8:2. The DPA3 model was fine-tuned for 200,000 training steps using an initial learning rate of $1.0 \times 10^{-4}$ and a final learning rate of $3.0 \times 10^{-8}$, with decay applied every 5,000 steps. The fine-tuned machine-learning potential was used within ASE to accelerate structural relaxation, energy evaluation, and phonon-based stability prescreening.\cite{ref59} For thermodynamic screening, candidate structures were relaxed using BFGS as the primary optimizer, while FIRE was adopted as a fallback optimizer when needed. Relative thermodynamic stability was then evaluated using the energy above the convex hull, where competing phases were obtained from experimentally validated compounds in the corresponding chemical spaces of the Materials Project database.\cite{ref60} For dynamical screening, phonon calculations were performed at the machine-learning-potential level within ASE using the finite-displacement method. Structures were considered dynamically stable when the minimum phonon frequency exceeded -0.1 THz.

\subsection{First-principles calculations and property evaluation}

High-throughput first-principles calculations were carried out within the JAMIP (Jilin Artificial-intelligence aided Materials-design Integrated Package) automation framework\cite{ref61} with electronic-structure calculations performed using VASP.\cite{ref62} Electron-ion interactions were described using the projector augmented-wave (PAW) method, and exchange-correlation effects were treated within the Perdew-Burke-Ernzerhof (PBE) generalized gradient approximation. For transition-metal-containing systems, the GGA+U approach was adopted to account for localized d-electron correlations. All crystal structures were fully optimized using a plane-wave cutoff energy of 500 eV and a k-point spacing of \SI{0.03}{\per\angstrom}. The convergence criteria were set to $10^{-5}$ eV for total energy and \SI{0.01}{\electronvolt\per\angstrom} for atomic forces. For selected candidates that passed the stability and electronic-structure screening, DFT phonon spectra were further calculated using Phonopy with the finite-displacement method.\cite{ref63,ref64} Supercells were constructed such that all lattice dimensions after expansion exceeded \SI{10}{\angstrom}. For key candidates, band gaps and band structures were further evaluated using the HSE06 screened hybrid functional with 25\% Hartree-Fock exchange. Optical absorption coefficients were obtained from the frequency-dependent dielectric function calculated at the PBE level and corrected using a scissor operator based on the HSE06 band gap. Based on this corrected data, the spectral limited maximum efficiency (SLME) was evaluated. Thermoelectric properties were initially calculated using PBE-level band structures within BoltzTraP2\cite{ref65} under the constant-relaxation-time approximation at 300 K to obtain Seebeck coefficients and the electronic fitness function (EFF) as functions of carrier concentration.\cite{ref48} Following the initial thermoelectric screening, the selected candidates were further examined using the modified Becke-Johnson (mBJ) potential to refine the thermoelectric-property evaluation.

\begin{acknowledgement}

This work was supported by the National Natural Science Foundation of China (Grants No.62125402 and 62321166653) and Jilin Province Science and Technology Development Program (Grant No. 20240101312JC). Calculations were performed in part at the High Performance Computing Center of Jilin University.

\end{acknowledgement}

\bibliography{references}

\clearpage
\section*{Supporting Information}
\renewcommand{\figurename}{Supplementary Figure}
\renewcommand{\tablename}{Supplementary Table}
\setcounter{figure}{0}
\setcounter{table}{0}

\subsection*{Robust AI-Driven Discovery of Electronic Metal Phosphide Semiconductors}

Benhao Zhu\textsuperscript{1,\#}, Muhammad Faizan\textsuperscript{1,\#}, Zewei Li\textsuperscript{1}, Wenshuo Li\textsuperscript{1}, Feifei Ren\textsuperscript{1}, Jiahao Xie\textsuperscript{1,*}, Lijun Zhang\textsuperscript{1,*}

\textsuperscript{1}State Key Laboratory of Integrated Optoelectronics, Key Laboratory of Automobile Materials and Key Laboratory of Material Simulation Methods and Software of MOE, School of Materials Science and Engineering, Jilin University, Changchun, Jilin 130012, China

\textsuperscript{\#} These authors contributed equally to this work

*Corresponding authors: xiejh.mail@gmail.com; lijun\_zhang@jlu.edu.cn.


\begingroup
\small
\setlength{\tabcolsep}{3pt}
\renewcommand{\arraystretch}{1.16}
\begin{longtable}[]{@{}
  >{\centering\arraybackslash}p{(\linewidth - 8\tabcolsep) * \real{0.1769}}
  >{\centering\arraybackslash}p{(\linewidth - 8\tabcolsep) * \real{0.2991}}
  >{\centering\arraybackslash}p{(\linewidth - 8\tabcolsep) * \real{0.1614}}
  >{\centering\arraybackslash}p{(\linewidth - 8\tabcolsep) * \real{0.1723}}
  >{\centering\arraybackslash}p{(\linewidth - 8\tabcolsep) * \real{0.1903}}@{}}
\caption{Space group, prototype formula, and number of atoms per primitive cell ($N_\mathrm{pc}$) for the 20 structural prototypes derived from phosphide entries in the ICSD through Wyckoff-site analysis.}\label{tab:si-1}\\
\toprule\noalign{}
\begin{minipage}[b]{\linewidth}\centering
Representation
\end{minipage} & \begin{minipage}[b]{\linewidth}\centering
Prototype label
\end{minipage} & \begin{minipage}[b]{\linewidth}\centering
$N_\mathrm{pc}$
\end{minipage} & \begin{minipage}[b]{\linewidth}\centering
Space Group
\end{minipage} & \begin{minipage}[b]{\linewidth}\centering
Prototype Formulas
\end{minipage} \\
\midrule\noalign{}
\endfirsthead
\caption[]{Space group, prototype formula, and number of atoms per primitive cell ($N_\mathrm{pc}$) for the 20 structural prototypes derived from phosphide entries in the ICSD through Wyckoff-site analysis. (continued)}\\
\toprule\noalign{}
\begin{minipage}[b]{\linewidth}\centering
Representation
\end{minipage} & \begin{minipage}[b]{\linewidth}\centering
Prototype label
\end{minipage} & \begin{minipage}[b]{\linewidth}\centering
$N_\mathrm{pc}$
\end{minipage} & \begin{minipage}[b]{\linewidth}\centering
Space Group
\end{minipage} & \begin{minipage}[b]{\linewidth}\centering
Prototype Formulas
\end{minipage} \\
\midrule\noalign{}

\endhead
\bottomrule\noalign{}
\endlastfoot
ICSD-12145 & 139\_2a\_4d\_4e & 10 & \(I4/mmm\) & $XY_2P_2$ \\
ICSD-157629 & 174\_6j6k\_a3j3k\_cf & 21 & \(P\overline{6}\) & $X_2Y_{12}P_7$ \\
ICSD-49728 & 62\_4c\_4c\_4c & 12 & \(Pnma\) & XYP \\
ICSD-65715 & 62\_12c\_20c\_4c & 36 & \(Pnma\) & $X_5YP_3$ \\
ICSD-41706 & 194\_2a\_2c\_2d & 6 & \(P6_{3}/mmc\) & XYP \\
ICSD-42016 & 164\_2d\_2d\_a & 5 & \(P\overline{3}m1\) & $XY_2P_2$ \\
ICSD-402227 & 189\_3f\_3g\_b2c & 9 & \(P\overline{6}2m\) & XYP \\
ICSD-245293 & 204\_24g\_2a\_8c & 34 & \(Im\overline{3}\) & $X_1Y_4P_{12}$ \\
ICSD-121379 & 122\_4a\_4b\_8d & 16 & \(I\overline{4}2d\) & $XYP_2$ \\
ICSD-68525 & 136\_2b\_4g\_8i & 14 & \(P4_{2}/mnm\) & $X_1Y_4P_2$ \\
ICSD-624057 & 59\_2b4e\_4a2b12e\_4a8e & 36 & \(Pmmn\) & $X_1Y_3P_2$ \\
ICSD-56444 & 187\_a\_d\_f & 3 & \(P\overline{6}m2\) & $X_1Y_1P_1$ \\
ICSD-42037 & 129\_2a\_2c\_2c & 6 & \(P4/nmm\) & XYP \\
ICSD-423780 & 15\_16f\_4e8f\_8f & 36 & \(C2/c\) & $X_3Y_4P_4$ \\
ICSD-67262 & 166\_12c\_3a\_6c & 21 & \(R\overline{3}m\) & $X_4Y_1P_2$ \\
ICSD-60077 & 14\_12e\_20e\_4e & 36 & \(P2_{1}/c\) & $X_5YP_3$ \\
ICSD-41181 & 14\_12e\_16e\_8e & 36 & \(P2_{1}/c\) & $X_3Y_2P_4$ \\
ICSD-60125 & 62\_12c\_12c\_4c & 28 & Pnma & $X_3YP_3$ \\
ICSD-16294 & 166\_18h\_6c & 24 & \(R\overline{3}m\) & $XP_3$ \\
ICSD-106350 & 221\_3c\_a\_b & 5 & \(Pm\overline{3}m\) & $X_3YP$ \\
\end{longtable}
\endgroup

\clearpage

\begin{landscape}
\begingroup
\scriptsize
\setlength{\tabcolsep}{2pt}
\renewcommand{\arraystretch}{1.16}
\begin{longtable}[]{@{}
  >{\centering\arraybackslash}p{(\linewidth - 18\tabcolsep) * \real{0.1136}}
  >{\centering\arraybackslash}p{(\linewidth - 18\tabcolsep) * \real{0.1068}}
  >{\centering\arraybackslash}p{(\linewidth - 18\tabcolsep) * \real{0.1269}}
  >{\centering\arraybackslash}p{(\linewidth - 18\tabcolsep) * \real{0.0818}}
  >{\centering\arraybackslash}p{(\linewidth - 18\tabcolsep) * \real{0.0962}}
  >{\centering\arraybackslash}p{(\linewidth - 18\tabcolsep) * \real{0.0829}}
  >{\centering\arraybackslash}p{(\linewidth - 18\tabcolsep) * \real{0.1134}}
  >{\centering\arraybackslash}p{(\linewidth - 18\tabcolsep) * \real{0.0681}}
  >{\centering\arraybackslash}p{(\linewidth - 18\tabcolsep) * \real{0.1047}}
  >{\centering\arraybackslash}p{(\linewidth - 18\tabcolsep) * \real{0.1055}}@{}}
\caption{Summary of the electronic-structure descriptors, photovoltaic performance, carrier effective masses, and crystal-structure parameters of the newly designed photoactive candidate materials.}\label{tab:si-2}\\
\toprule\noalign{}
\begin{minipage}[b]{\linewidth}\centering
Formula
\end{minipage} & \begin{minipage}[b]{\linewidth}\centering
Type
\end{minipage} & \begin{minipage}[b]{\linewidth}\centering
$E_g^\mathrm{PBE}$ (eV)
\end{minipage} & \begin{minipage}[b]{\linewidth}\centering
$E_g^\mathrm{HSE}$ (eV)
\end{minipage} & \begin{minipage}[b]{\linewidth}\centering
SLME
\end{minipage} & \begin{minipage}[b]{\linewidth}\centering
$m_h^*/m_0$
\end{minipage} & \begin{minipage}[b]{\linewidth}\centering
$m_e^*/m_0$
\end{minipage} & \begin{minipage}[b]{\linewidth}\centering
$N_\mathrm{pc}$
\end{minipage} & \begin{minipage}[b]{\linewidth}\centering
Space group
\end{minipage} & \begin{minipage}[b]{\linewidth}\centering
Space group number
\end{minipage} \\
\midrule\noalign{}
\endfirsthead
\caption[]{Summary of the electronic-structure descriptors, photovoltaic performance, carrier effective masses, and crystal-structure parameters of the newly designed photoactive candidate materials. (continued)}\\
\toprule\noalign{}
\begin{minipage}[b]{\linewidth}\centering
Formula
\end{minipage} & \begin{minipage}[b]{\linewidth}\centering
Type
\end{minipage} & \begin{minipage}[b]{\linewidth}\centering
$E_g^\mathrm{PBE}$ (eV)
\end{minipage} & \begin{minipage}[b]{\linewidth}\centering
$E_g^\mathrm{HSE}$ (eV)
\end{minipage} & \begin{minipage}[b]{\linewidth}\centering
SLME
\end{minipage} & \begin{minipage}[b]{\linewidth}\centering
$m_h^*/m_0$
\end{minipage} & \begin{minipage}[b]{\linewidth}\centering
$m_e^*/m_0$
\end{minipage} & \begin{minipage}[b]{\linewidth}\centering
$N_\mathrm{pc}$
\end{minipage} & \begin{minipage}[b]{\linewidth}\centering
Space group
\end{minipage} & \begin{minipage}[b]{\linewidth}\centering
Space group number
\end{minipage} \\
\midrule\noalign{}

\endhead
\bottomrule\noalign{}
\endlastfoot
\ce{RbZnP} & Direct & 0.67 & 1.36 & 33.06 & 0.75 & 0.13 & 6 & \(P6_{3}/mmc\)
& 194 \\
\ce{Sr3GaP3} & Direct & 0.61 & 1.2 & 32.19 &
0.35 & 0.43 & 28 & \(Pnma\) & 62 \\
\ce{Ca5SiP3} & Direct & 0.48 & 0.99 & 31.7 &
0.71 & 0.43 & 36 & \(P2_{1}/c\) & 14 \\
\ce{KCdP} & Direct & 0.26 & 0.95 & 30.76 & 0.68 & 0.06 & 6 & \(P6_{3}/mmc\) &
194 \\
\ce{Ca3BiP} & Direct & 0.71 & 1.2 & 30.56 & 0.37 & 0.43 & 10
& \(Pmmn\) & 59 \\
\ce{KSnP} & Direct & 0.8 & 1.36 & 30.52 & 0.49 & 0.14 & 3 & \(P3m1\) & 156 \\
\ce{Sr3BiP} & Indirect & 0.73 & 1.23 & 30.07 & 0.68 & 0.14 &
10 & \(Pmn2_{1}\) & 31 \\
\end{longtable}
\endgroup
\end{landscape}
\clearpage

\begin{landscape}
\begingroup
\scriptsize
\setlength{\tabcolsep}{2pt}
\renewcommand{\arraystretch}{1.16}
\begin{longtable}[]{@{}
  >{\centering\arraybackslash}p{(\linewidth - 20\tabcolsep) * \real{0.1363}}
  >{\centering\arraybackslash}p{(\linewidth - 20\tabcolsep) * \real{0.1033}}
  >{\centering\arraybackslash}p{(\linewidth - 20\tabcolsep) * \real{0.1026}}
  >{\centering\arraybackslash}p{(\linewidth - 20\tabcolsep) * \real{0.0684}}
  >{\centering\arraybackslash}p{(\linewidth - 20\tabcolsep) * \real{0.0684}}
  >{\centering\arraybackslash}p{(\linewidth - 20\tabcolsep) * \real{0.0856}}
  >{\centering\arraybackslash}p{(\linewidth - 20\tabcolsep) * \real{0.0684}}
  >{\centering\arraybackslash}p{(\linewidth - 20\tabcolsep) * \real{0.1027}}
  >{\centering\arraybackslash}p{(\linewidth - 20\tabcolsep) * \real{0.0507}}
  >{\centering\arraybackslash}p{(\linewidth - 20\tabcolsep) * \real{0.1197}}
  >{\centering\arraybackslash}p{(\linewidth - 20\tabcolsep) * \real{0.0939}}@{}}
\caption{Summary of the electronic-structure descriptors, photovoltaic performance, carrier effective masses, and crystal-structure parameters of the photoactive candidate materials identified from phosphide entries in the ICSD database.}\label{tab:si-3}\\
\toprule\noalign{}
\begin{minipage}[b]{\linewidth}\centering
Struct ID
\end{minipage} & \begin{minipage}[b]{\linewidth}\centering
Formula
\end{minipage} & \begin{minipage}[b]{\linewidth}\centering
Type
\end{minipage} & \begin{minipage}[b]{\linewidth}\centering
$E_g^\mathrm{PBE}$
\end{minipage} & \begin{minipage}[b]{\linewidth}\centering
$E_g^\mathrm{HSE}$
\end{minipage} & \begin{minipage}[b]{\linewidth}\centering
SLME
\end{minipage} & \begin{minipage}[b]{\linewidth}\centering
$m_h^*/m_0$
\end{minipage} & \begin{minipage}[b]{\linewidth}\centering
$m_e^*/m_0$
\end{minipage} & \begin{minipage}[b]{\linewidth}\centering
$N_\mathrm{pc}$
\end{minipage} & \begin{minipage}[b]{\linewidth}\centering
Space group
\end{minipage} & \begin{minipage}[b]{\linewidth}\centering
Space group number
\end{minipage} \\
\midrule\noalign{}
\endfirsthead
\caption[]{Summary of the electronic-structure descriptors, photovoltaic performance, carrier effective masses, and crystal-structure parameters of the photoactive candidate materials identified from phosphide entries in the ICSD database. (continued)}\\
\toprule\noalign{}
\begin{minipage}[b]{\linewidth}\centering
Struct ID
\end{minipage} & \begin{minipage}[b]{\linewidth}\centering
Formula
\end{minipage} & \begin{minipage}[b]{\linewidth}\centering
Type
\end{minipage} & \begin{minipage}[b]{\linewidth}\centering
$E_g^\mathrm{PBE}$
\end{minipage} & \begin{minipage}[b]{\linewidth}\centering
$E_g^\mathrm{HSE}$
\end{minipage} & \begin{minipage}[b]{\linewidth}\centering
SLME
\end{minipage} & \begin{minipage}[b]{\linewidth}\centering
$m_h^*/m_0$
\end{minipage} & \begin{minipage}[b]{\linewidth}\centering
$m_e^*/m_0$
\end{minipage} & \begin{minipage}[b]{\linewidth}\centering
$N_\mathrm{pc}$
\end{minipage} & \begin{minipage}[b]{\linewidth}\centering
Space group
\end{minipage} & \begin{minipage}[b]{\linewidth}\centering
Space group number
\end{minipage} \\
\midrule\noalign{}

\endhead
\bottomrule\noalign{}
\endlastfoot
ICSD-56444 & \ce{BaLiP} & Direct & 0.70 & 1.34 & 33.41 & 0.22 & 0.19 & 3 &
\(P\overline{6}m2\) & 187 \\
ICSD-35342 & \ce{Ba3SnP3} & Indirect & 0.84 &
1.41 & 33.15 & 0.67 & 0.77 & 28 & \(P2_{1}/c\) & 14 \\
ICSD-1153 & \ce{Na2CuP} & Direct & 0.91 & 1.35 & 32.93 & 0.63
& 0.30 & 16 & \(Cmcm\) & 63 \\
ICSD-48163 & \ce{P2Pd} & Direct & 0.69 & 1.27 & 32.91 & 0.45
& 0.28 & 12 & \(C2/c\) & 15 \\
ICSD-30915 & \ce{Ba(CdP)2} & Indirect & 0.69 & 1.29 & 32.75 &
0.54 & 0.15 & 5 & \(P\overline{3}m1\) & 164 \\
ICSD-300146 & \ce{Na3InP2} & Direct & 0.65 &
1.31 & 32.74 & 0.27 & 0.10 & 48 & \(P2_{1}/c\) & 14 \\
ICSD-429732 & \ce{BaP2} & Direct & 0.57 & 1.16 & 32.44 & 0.70
& 0.55 & 18 & \(P2_{1}/c\) & 14 \\
ICSD-61335 & \ce{Sr3(InP2)2}
& Direct & 0.59 & 1.29 & 32.12 & 0.26 & 0.12 & 18 & \(Pnnm\) & 58 \\
ICSD-65185 & \ce{Cs2P3} & Direct & 0.84 &
1.43 & 32.03 & 0.83 & 0.59 & 40 & \(Fmmm\) & 69 \\
ICSD-35283 & \ce{AgP2} & Indirect & 0.70 & 1.33 & 31.99 &
0.34 & 0.49 & 12 & \(P2_{1}/c\) & 14 \\
ICSD-198284 & \ce{InP} & Direct & 0.47 & 1.24 & 31.95 & 0.46 & 0.05 & 8 &
\(F\overline{4}3m\) & 216 \\
ICSD-33259 & \ce{K2P3} & Direct & 0.70 & 1.32
& 31.86 & 0.70 & 0.58 & 40 & \(Fmmm\) & 69 \\
ICSD-250015 & \ce{ZnP2} & Direct & 0.75 & 1.45 & 31.77 & 0.39
& 0.28 & 24 & \(P2_{1}/c\) & 14 \\
ICSD-65056 & \ce{Sr3InP3} & Direct & 0.84 &
1.48 & 31.76 & 0.35 & 0.38 & 28 & \(Pnma\) & 62 \\
ICSD-402227 & BaNaP & Direct & 0.86 & 1.52 & 31.28 & 0.32 & 0.27 & 9 &
\(P\overline{6}2m\) & 189 \\
ICSD-27160 & \ce{NiP2} & Indirect & 0.60 & 1.28 & 31.10 &
0.29 & 0.77 & 12 & \(C2/c\) & 15 \\
ICSD-25605 & \ce{CdP4} & Indirect & 0.60 & 1.19 & 31.09 &
0.21 & 0.45 & 10 & \(P2_{1}/c\) & 14 \\
ICSD-91560 & \ce{NiP2} & Indirect & 0.59 & 1.31 & 31.02 &
0.28 & 0.78 & 12 & \(C2/c\) & 15 \\
ICSD-402811 & \ce{Rb3InP2} & Indirect & 0.49
& 1.06 & 30.94 & 0.93 & 0.17 & 48 & \(P\overline{1}\) & 2 \\
ICSD-416890 & \ce{BaLiP} & Direct & 0.96 & 1.58 & 30.83 & 0.56 & 0.25 & 6 &
\(P6_{3}/mmc\) & 194 \\
ICSD-22179 & \ce{ZnSnP2} & Direct & 0.70 & 1.53 & 30.82 &
0.25 & 0.06 & 16 & \(I\overline{4}2d\) & 122 \\
ICSD-42015 & \ce{LaP2} & Direct & 0.46 & 1.05 & 30.75 & 0.56
& 0.59 & 48 & \(Cc\) & 9 \\
ICSD-22183 & \ce{CdSnP2} & Direct & 0.27 & 1.02 & 30.40 &
0.08 & 0.04 & 16 & \(I\overline{4}2d\) & 122 \\
\end{longtable}
\endgroup
\end{landscape}
\clearpage

\begin{figure}[!p]
\centering
\includegraphics[height=0.58\textheight,width=\linewidth,keepaspectratio]{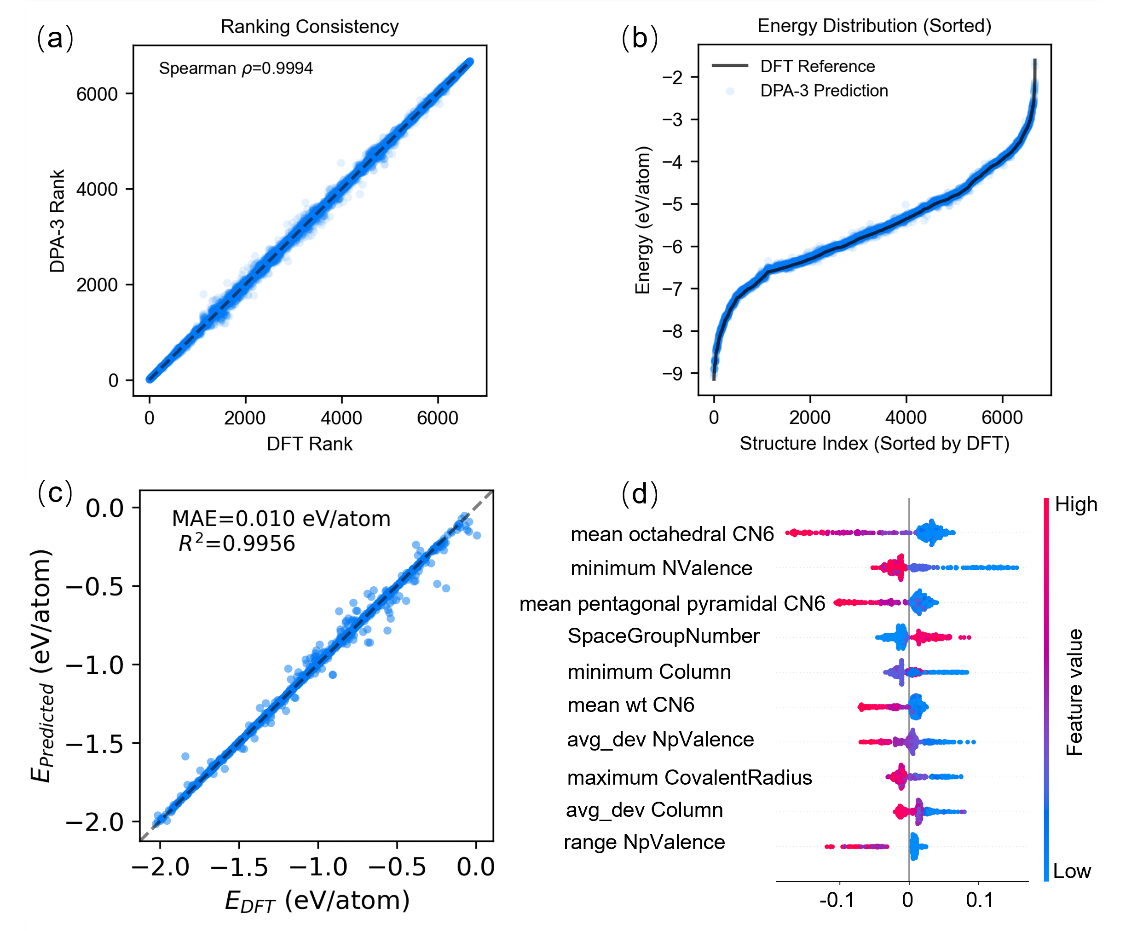}
\caption{(a) Parity plot of per-atom energy ranking predicted by the machine-learning potential, (b) parity plot of formation-energy prediction by the random-forest model trained on handcrafted descriptors, and SHAP analysis of feature importance for the random-forest model.}
\label{fig:si-mlp-parity}
\end{figure}
\FloatBarrier

\begin{figure}[!p]
\centering
\includegraphics[height=0.84\textheight,width=\linewidth,keepaspectratio]{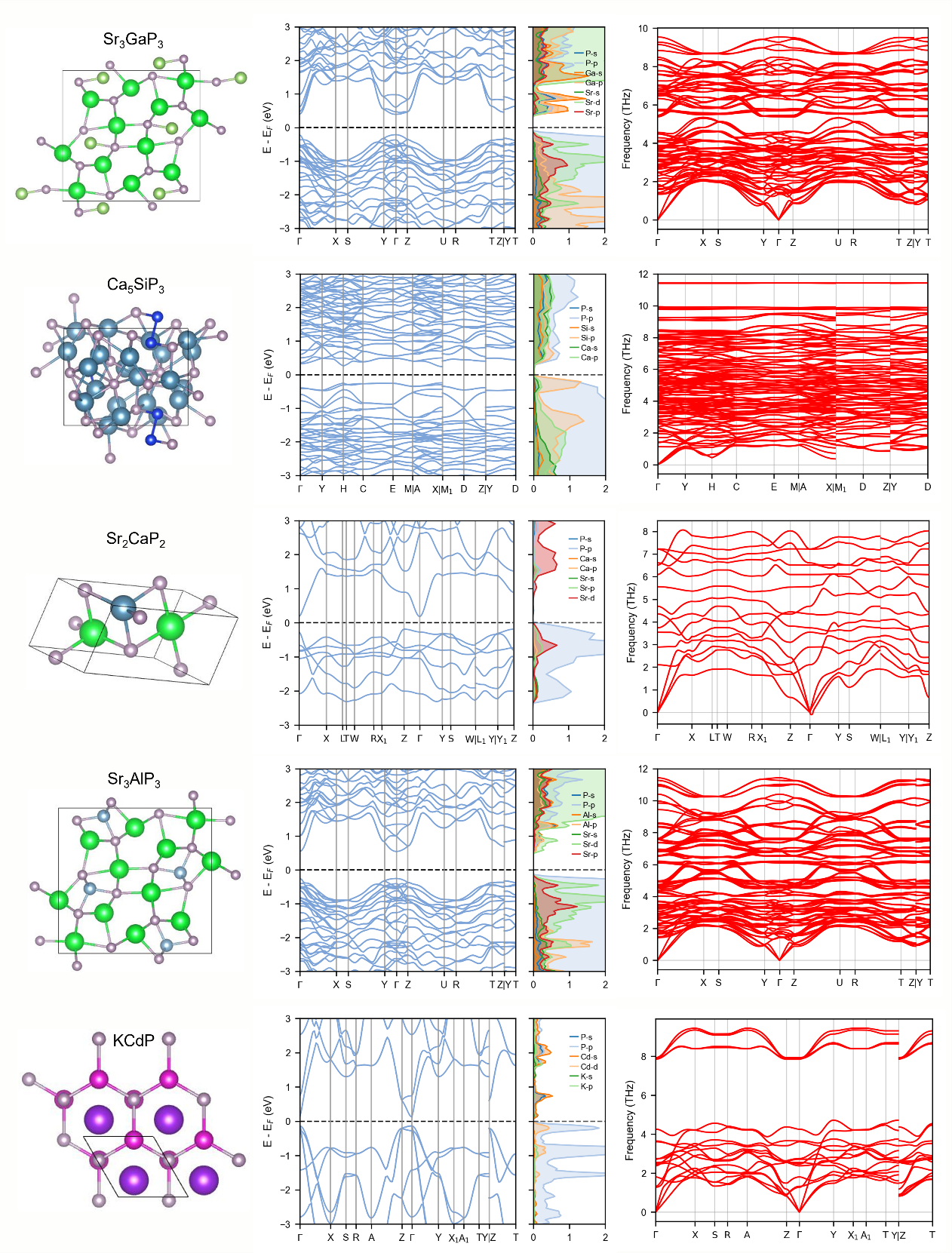}
\end{figure}
\FloatBarrier

\begin{figure}[!p]
\centering
\includegraphics[height=0.84\textheight,width=\linewidth,keepaspectratio]{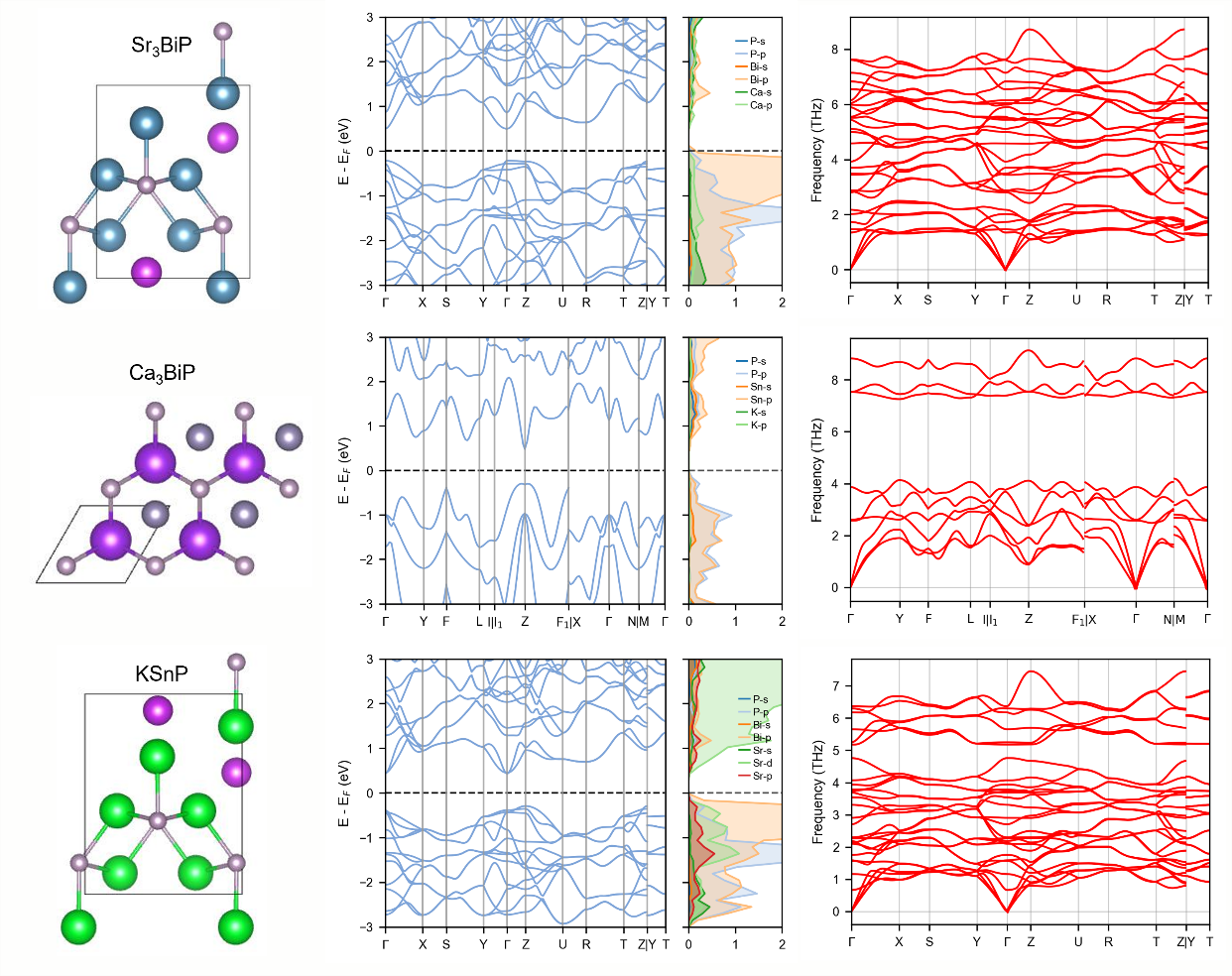}
\caption{Crystal structures, band structures, density of states, and phonon spectra of the newly designed photoactive candidate materials.}
\label{fig:si-new-photoactive}
\end{figure}
\FloatBarrier

\begin{figure}[!p]
\centering
\includegraphics[height=0.38\textheight,width=\linewidth,keepaspectratio]{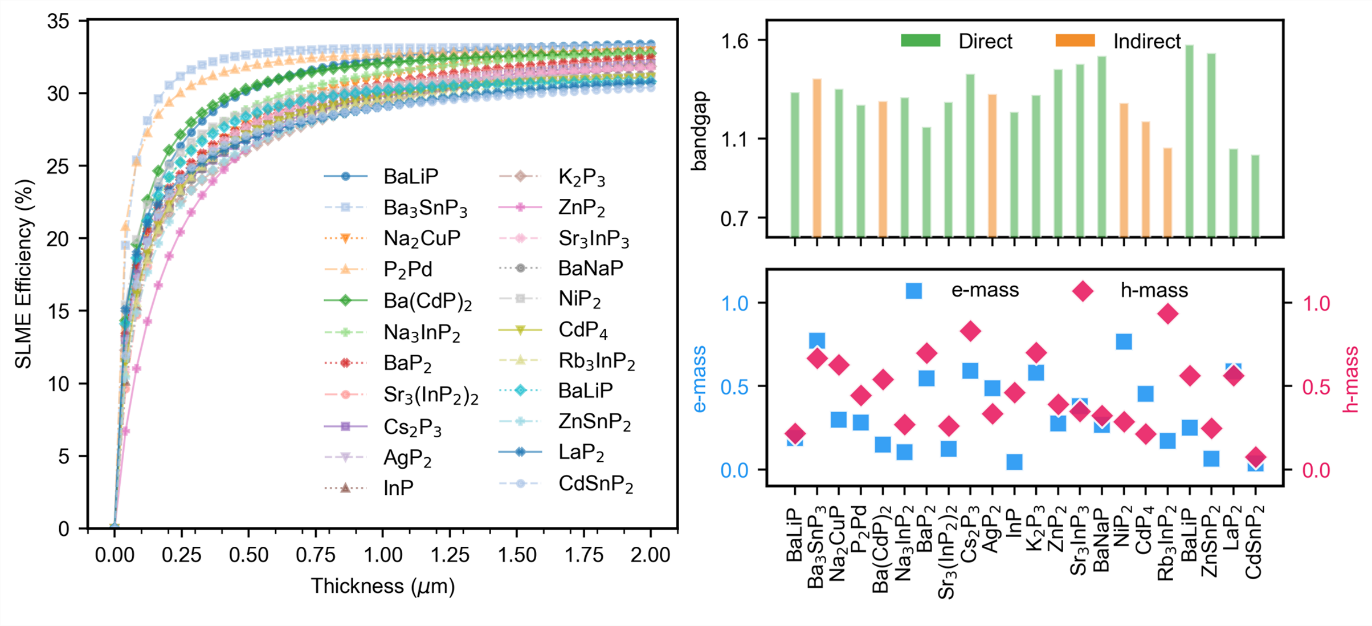}
\caption{SLME as a function of absorber thickness, together with the HSE06 band gaps and carrier effective masses for individual photoactive candidate materials identified from phosphide entries in the ICSD database.}
\label{fig:si-known-photoactive-slme}
\end{figure}
\FloatBarrier

\begin{figure}[!p]
\centering
\includegraphics[height=0.82\textheight,width=\linewidth,keepaspectratio]{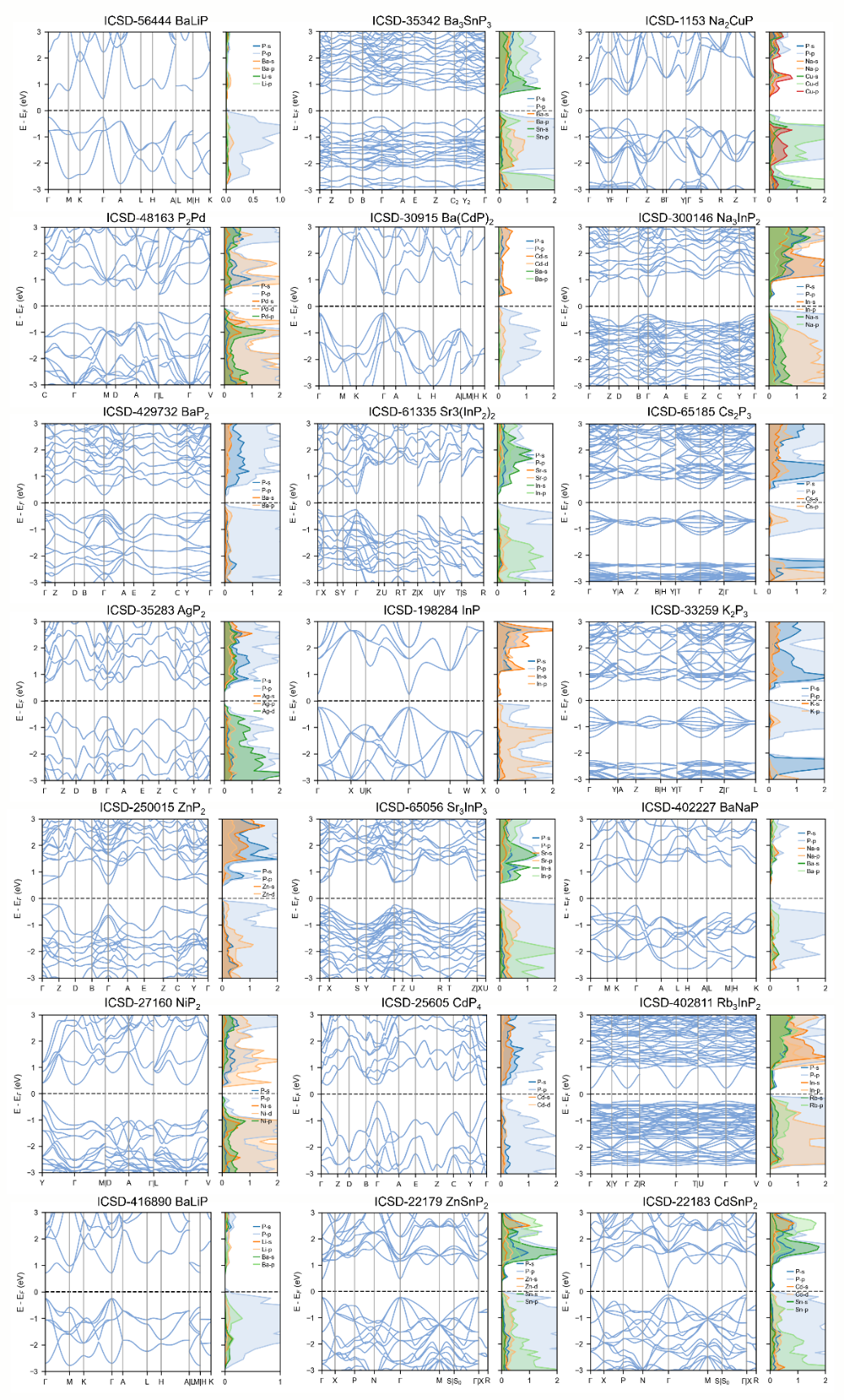}
\caption{Band structures and density of states of representative photoactive candidate materials identified from phosphide entries in the ICSD database}
\label{fig:si-known-photoactive-bands}
\end{figure}
\FloatBarrier

\begin{figure}[!p]
\centering
\includegraphics[height=0.84\textheight,width=\linewidth,keepaspectratio]{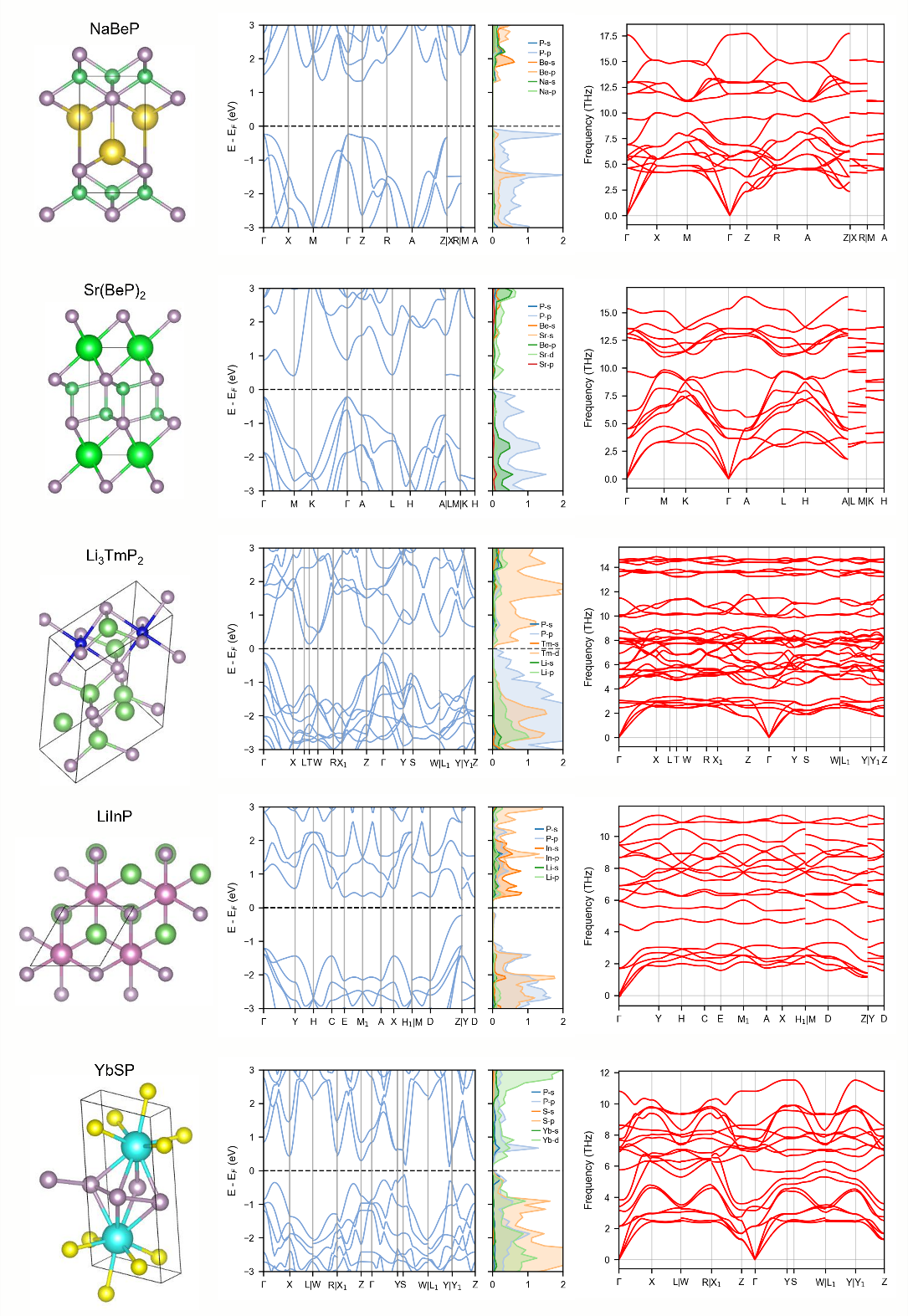}
\end{figure}
\FloatBarrier

\begin{figure}[!p]
\centering
\includegraphics[height=0.84\textheight,width=\linewidth,keepaspectratio]{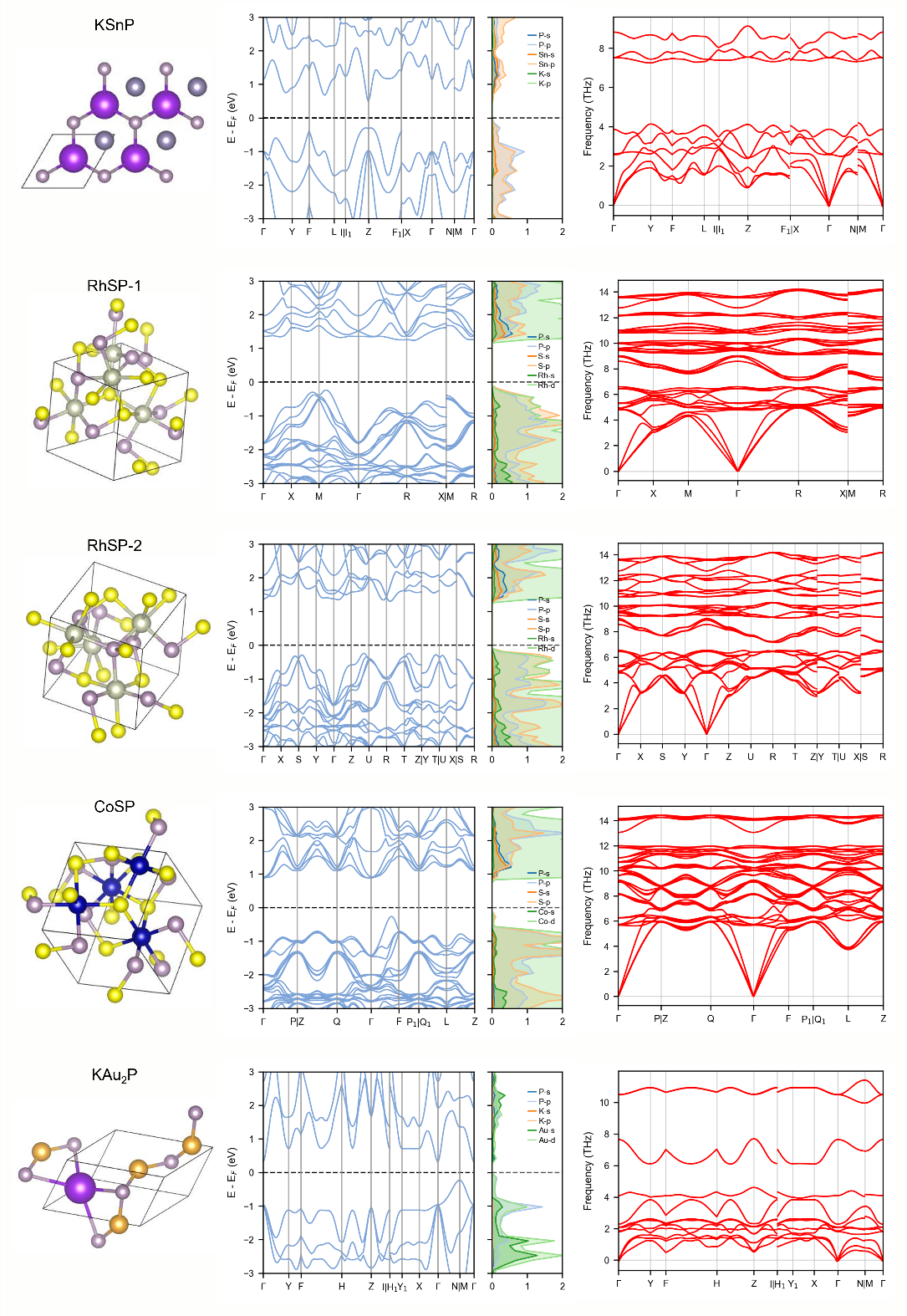}
\caption{Crystal structures, band structures, density of states, and phonon spectra of newly designed thermoelectric candidate materials.}
\label{fig:si-new-thermoelectric}
\end{figure}
\FloatBarrier

\begin{figure}[!p]
\centering
\includegraphics[height=0.58\textheight,width=\linewidth,keepaspectratio]{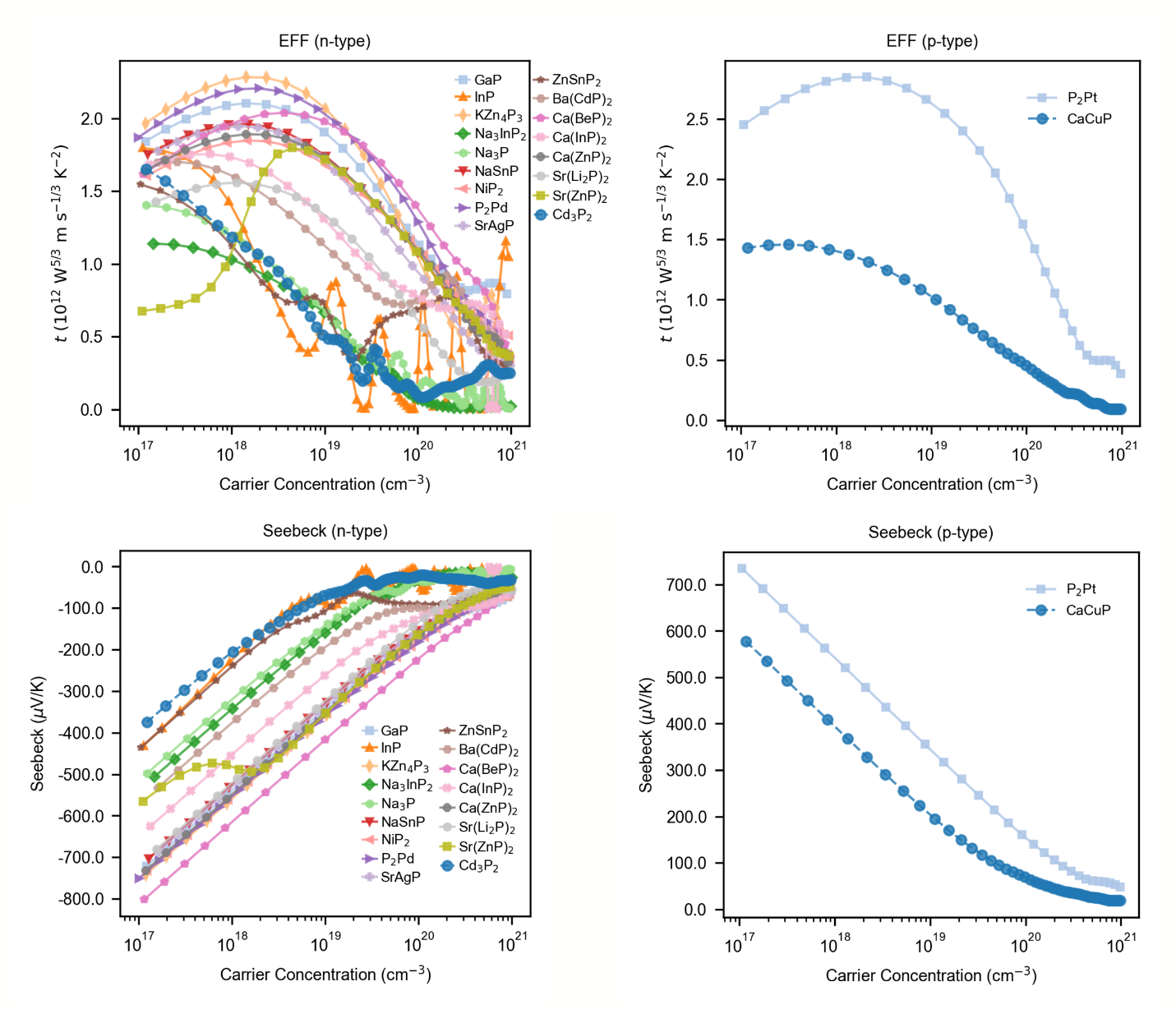}
\caption{Electronic fitness function (EFF) and Seebeck coefficient as functions of carrier concentration for the thermoelectric candidate materials identified from phosphide entries in the ICSD database.}
\label{fig:si-known-te-eff}
\end{figure}
\FloatBarrier

\begin{figure}[!p]
\centering
\includegraphics[height=0.82\textheight,width=\linewidth,keepaspectratio]{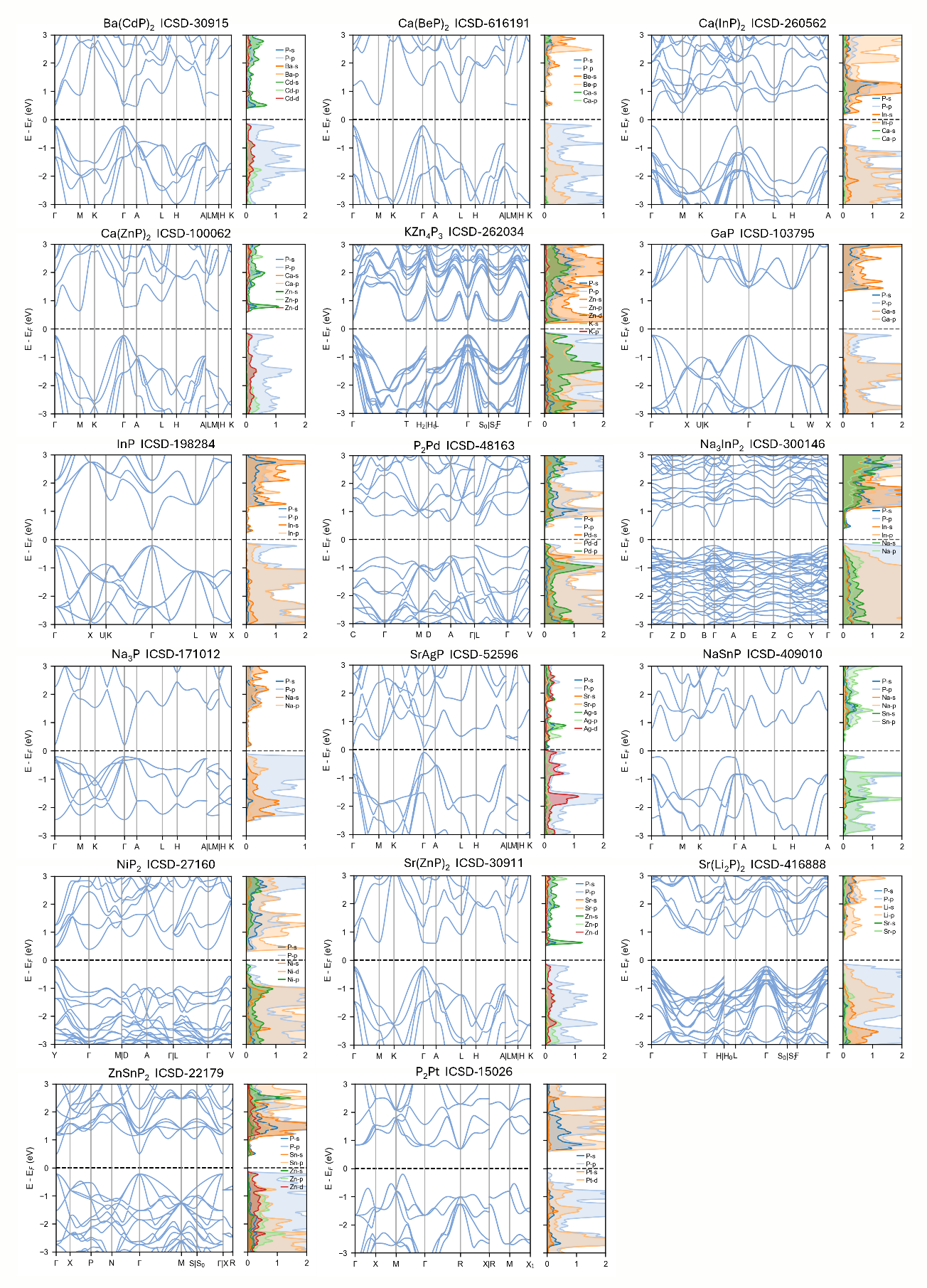}
\caption{PBE band structures and density of states of the thermoelectric candidate materials identified from phosphide entries in the ICSD database.}
\label{fig:si-known-te-bands}
\end{figure}
\FloatBarrier

\begin{figure}[!p]
\centering
\includegraphics[height=0.38\textheight,width=\linewidth,keepaspectratio]{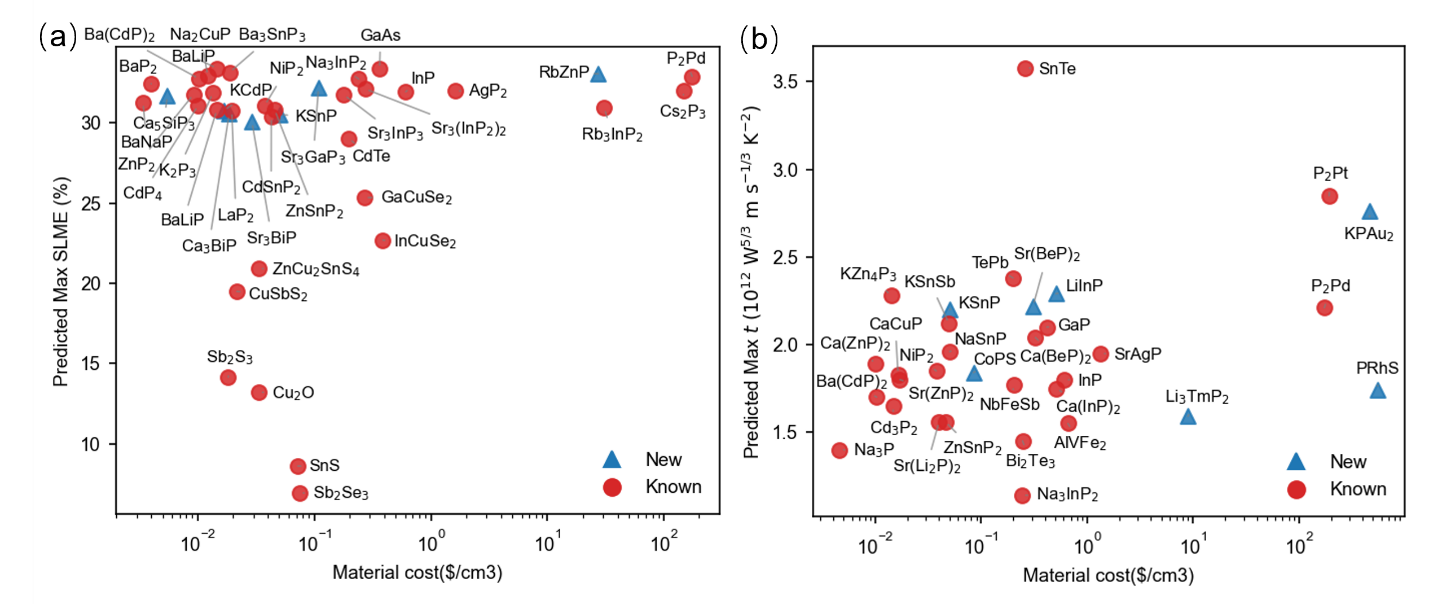}
\caption{Relationship between material cost and performance metrics for newly designed and known compounds. Blue triangles and red circles denote newly designed and known compounds, respectively. (a) Material cost versus SLME at a thickness of \SI{2}{\micro\meter} for screened optoelectronic candidates. (b) Material cost versus peak thermoelectric efficiency (EFF) for screened thermoelectric candidates.}
\label{fig:si-cost-performance}
\end{figure}
\FloatBarrier

\end{document}